\documentclass[12pt]{article}
\usepackage{amsmath}
\usepackage{amsfonts}
\usepackage{amsmath}
\usepackage{latexsym}
\usepackage{amsfonts}
\usepackage{graphics}
\usepackage{graphicx}
\graphicspath{ {./images/} }
\usepackage{jstyle}

\newcommand*{\Scale}[2][4]{\scalebox{#1}{$#2$}}

\newcommand\gf[2]{\left(\genfrac{}{}{0pt}{}{#1}{#2}\right)}
\newcommand\gff[2]{\left[\genfrac{}{}{0pt}{}{#1}{#2}\right]}
\usepackage{tikz}
\usetikzlibrary{arrows}

\title{On correlation functions of higher-spin currents in arbitrary dimensions $d>3$}

\author[a]{Melik Karapetyan}
\author[a]{Ruben Manvelyan}
\author[b]{Karapet Mkrtchyan}

\affiliation[a]{A. Alikhanyan National Laboratory (Yerevan Physics Institute) Alikhanian Br. Str. 2, 0036 Yerevan, Armenia}
\affiliation[b]{Theoretical Physics Group, Blackett Laboratory, Imperial College London SW7 2AZ, U.K.}

\emailAdd{meliq.karapetyan@gmail.com}
\emailAdd{manvel@yerphi.am}
\emailAdd{k.mkrtchyan@imperial.ac.uk}

\abstract{We revisit the problem of classification and explicit construction of the conformal three-point correlation functions of currents of arbitrary integer spin in arbitrary dimensions. For the conserved currents, we set up the equations for the conservation conditions and solve them completely for some values of spins, confirming the earlier counting of the number of independent structures matching them with the higher-spin cubic vertices in one higher dimension. The general solution for the correlators of conserved currents we delegate to a follow-up work.}

\def \ollr{{\raise7pt\hbox{$\leftrightarrow  \! \! \! \! \! \!$}}}
\def \r{\rangle}
\def \de{\delta}
\def \si{\sigma}

\def \pr{\partial}
\def \hX{{\hat X}}

\def \hx{{\hat x}}

\def \ta{{\tilde a}}
\def \tb{{\tilde b}}
\def \tc{{\tilde c}}
\def \tC{{\tilde C}}

\def \tH{{\tilde H}}

\def \A{{\cal A}}

\def \E{{\cal E}}

\def \I{{\cal I}}
\def \J{{\cal J}}
\def \O{{\cal O}}

\begin{document}

{\phantom{.}\vspace{-2.5cm}\\\flushright Imperial-TP-KM-2023-02\\ }

\maketitle



\section{Introduction}

\quad 

The holographic duality \cite{Maldacena:1997re,Witten:1998qj} remains one of the most promising approaches to Quantum Gravity. Particular interest is attracted by Higher-Spin (HS) Gravity \cite{Vasiliev:1990en,Vasiliev:2003ev,Didenko:2014dwa} as the AdS dual candidate \cite{Sezgin:2002rt,Klebanov:2002ja,Giombi:2016ejx,Giombi:2016pvg} of the simplest CFT --- $O(N)$ vector model \cite{Lang:1991kp,Lang:1992pp}. Lagrangian formulation of Vasiliev's HS Gravity is not available so far. However, the classification of interaction vertices between symmetric HS fields in arbitrary dimensions has been an impressive collective effort. See \cite{Bengtsson:1983pd,Berends:1984wp,Berends:1984rq,Fradkin:1987ks,Fradkin:1986qy,Bengtsson:1986kh,Fradkin:1991iy,Metsaev:1991mt,Metsaev:1991nb,Vasiliev:2001wa,Alkalaev:2002rq,Manvelyan:2004mb,Bekaert:2005jf,Metsaev:2005ar,Bekaert:2006us,Boulanger:2006gr,Francia:2007qt,Fotopoulos:2007yq,Metsaev:2007rn,Fotopoulos:2008ka,Zinoviev:2008ck,Boulanger:2008tg,Manvelyan:2009tf,Manvelyan:2009vy,Bekaert:2009ud,Manvelyan:2010wp,Manvelyan:2010jr,Sagnotti:2010at,Zinoviev:2010cr,Fotopoulos:2010ay,Manvelyan:2010je,Polyakov:2010sk,Ruehl:2011tk,Vasiliev:2011knf,Joung:2011ww,Dempster:2012vw,Joung:2012rv,Buchbinder:2012iz,Henneaux:2012wg,Joung:2012fv,Manvelyan:2012ww,Joung:2012hz,Boulanger:2012dx,Henneaux:2013gba,Joung:2013nma,Conde:2016izb,Bengtsson:2016hss,Francia:2016weg,Taronna:2017wbx,Roiban:2017iqg,Sleight:2017pcz,Sleight:2017cax,Karapetyan:2019psg,Joung:2019wbl,Fredenhagen:2019lsz,Khabarov:2020bgr,Karapetyan:2021wdc} for some key references.

The holographic dictionary relates interaction vertices in AdS space-time to the conformal correlators on the boundary. Massless HS fields in AdS correspond to conserved currents on the boundary. The classification of the correlators of the (conserved) currents of arbitrary spin has been an independent parallel program. See \cite{Polyakov:1970xd,Schreier:1971um,Migdal:1971fof,Ferrara:1973yt,Ruehl:1973nj,Ruehl:1973pr,Koller:1974ut,Mack:1976pa,Osborn:1993cr,Osborn:1994rv,Erdmenger:1996yc,Park:1997bq,Osborn:1998qu,Park:1999pd,Anselmi:1999bb,Kuzenko:1999pi,Park:1999cw,Giombi:2009wh,Giombi:2010vg,Giombi:2011rz,Costa:2011mg,Costa:2011dw,Maldacena:2011jn,Stanev:2012nq,Todorov:2012xx,Zhiboedov:2012bm,Alba:2013yda,Costa:2014rya,Alba:2015upa,Kravchuk:2016qvl,Skvortsov:2018uru,Buchbinder:2022kmj,Buchbinder:2022mys,Buchbinder:2023coi} for some key references.

Generally, in conformal field theory, two and three-point correlation functions are fixed by conformal symmetry leaving no functional freedom. While the two-point function is fixed up to a normalization constant for any spin conformal operator (or traceless current of any rank) the three-point function depends on several constants for each triplet of currents. It is natural to expect that the number of independent structures here should match the number of independent vertices of cubic interaction in the bulk AdS gravity, via AdS/CFT dictionary. Moreover, the cubic vertices in AdS are uniquely determined from the flat space cubic vertices, by adding curvature corrections fixed by the requirement of AdS covariance \cite{Zinoviev:2008ck,Boulanger:2008tg,Manvelyan:2009tf,Vasiliev:2011knf,Joung:2011ww,Manvelyan:2012ww,Joung:2013nma,Francia:2016weg}. 
Hence, there should be a one-to-one correspondence between cubic vertices in $d+1$-dimensional Minkowski space and conformal corellators in $d$ dimensions. At least, the number of structures on both sides should match. This one-to-one correspondence between three-point correlators of conserved currents of arbitrary spin in $d>3$ dimensions and cubic vertices of massless symmetric fields in $d+1$ dimensional Minkowski space \cite{Metsaev:2005ar,Manvelyan:2010jr} was conjectured and elaborated upon in \cite{Costa:2011mg} (see also \cite{Costa:2011dw,Zhiboedov:2012bm}). 

Four-dimensional bulk spacetime corresponding to three-dimensional CFT has some peculiarities (see, e.g., \cite{Conde:2016izb,Giombi:2011rz,Skvortsov:2018uru}), while similar correspondence has been established in $d=2$ (with three-dimensional bulk) not only at cubic order but also for arbitrary higher-order interactions \cite{Fredenhagen:2018guf} with the help of the full classification of cubic \cite{Mkrtchyan:2017ixk,Kessel:2018ugi} and higher-order \cite{Fredenhagen:2019hvb} independent vertices involving massless bosonic HS fields.

The holographic reconstruction of HS Gravity has also progressed in the last decades: see \cite{Leonhardt:2003sn,Manvelyan:2005ew,Manvelyan:2006zy,Manvelyan:2005fp,Manvelyan:2008ks,Costa:2014kfa,Beccaria:2014xda,Bekaert:2014cea,Bekaert:2015tva,Sleight:2016dba,Ponomarev:2016jqk,Sleight:2016hyl,Ponomarev:2017qab} for some key references.

In this work, we revisit the construction and investigation of two and three-point correlation functions for HS conformal currents in arbitrary dimensions via Osborn-Petkou general formulation \cite{Osborn:1993cr}. In Appendix \ref{A} we briefly review this formulation adopted for higher spin case. But here we would like to note that the main advantage of formulation developed in \cite{Osborn:1993cr} is the reduction of the problem to construction instead of correlation function depending on three space-time points to the tensor depending on three sets of symmetrized indices but depending only on one variable which is roughly the difference of two coordinates inverted around the third point. In this way, we have a much simpler object for investigation depending on one variable polynomially with certain symmetry properties and satisfying conservation conditions. 

In this work, we present a general Ansatz for the local object that defines the correlation functions\footnote{We work with symmetric currents in arbitrary dimensions and do not consider lower-dimensional aspects like Schouten identities (relevant in $d\leq 3$) and parity-odd correlators (relevant in $d\leq 4$).} of arbitrary-spin currents. This Ansatz is a sum of the most general tensorial polynomials in {\it one} space-time variable and Kronecker symbols. Then we apply the symmetry conditions described in \cite{Osborn:1993cr} (see also Appendix \ref{A}) for general three-point correlation function with different spins $s_{1}, s_{2},s_{3}$. 
The ansatz we use here has a symmetry when exchanging the different currents of the same spin, differing from the more general ansatz of \cite{Costa:2011mg}. It is, however, general enough for the conserved currents, as the correlators of the latter are (anti)symmetric under the exchange of the currents of same spin, which is true also for the bulk vertices \cite{Manvelyan:2010jr}.
Natural triangle inequalities stem from the locality of our Ansatz. The solution of the latter is not simple, as expected (the approach of \cite{Osborn:1993cr} is known to lead to complications). However, we present the general solution in Section \ref{Ansatz}, reproducing all low spin examples presented in \cite{Osborn:1993cr}. The number of correlators of non-conserved currents (long representations) we count coincides with the results of \cite{Costa:2011mg} for non-coincident spins. However, the counting of correlators, (anti)symmetric under the exchange of the coincident spin currents, is new, as spelled out in detail in Section \ref{Ansatz}. The extrapolation of the general case would give a different number, counting all correlators, not only symmetric ones. Our new counting of ``symmetric correlators'', in particular, is relevant for coincident currents.

Then in the next section (Section $4$) we derive conservation conditions for our general ansatz. This allows investigation by computer calculation of the rank of an equivalent linear system of equations for getting independent parameters of the ansatz. One obtains general restriction on the number of independent parameters of the three-point function. Our results align with those of \cite{Costa:2011mg} (establishing one-to-one correspondence with the Minkowski vertices of massless fields \cite{Metsaev:2005ar,Manvelyan:2010jr}): {\it The number of independent parameters of the parity-even three-point function of three conserved currents depends only on the minimal spin of the involved currents and is equal to:} $\text{min}(s_1,s_2,s_3)+1.$

We further formulate the conservation condition in the form of a differential equation on the generating function of the correlators instead of a recursion relation for coefficients of the ansatz. We leave the full solution of these relations to future work.

Some technical details and derivations are delegated to Appendices.

\section{General setup and two-point function}
\renewcommand{\theequation}{\arabic{section}.\arabic{equation}}\setcounter{equation}{0}

\quad We present very shortly the key points of our technical setup and construction of the two-point function as a preliminary exercise  before our main task: the three-point function. As customary when dealing with HS fields, we introduce auxiliary vector variables $a_{\mu}, b_{\mu},\dots$  to handle an arbitrary number of symmetrized indices. As usual, we utilize instead of symmetric tensors such as $h^{(s)}_{\mu_1\mu_2...\mu_s}(x)$ the homogeneous polynomials in a vector $a^{\mu}$ of degree $s$ at the base point $x$:
\begin{equation}
h^{(s)}(x;a) = h^{(s)}_{\mu_1\mu_2...\mu_s}(x)a^{\mu_{1}}a^{\mu_{2}}\dots a^{\mu_{s}} .\label{0.1}
\end{equation}
Then the symmetrized gradient, divergence, and trace operations are given as\footnote{To distinguish easily between ``a'' and ``x'' spaces we introduce the notation $\nabla_{\mu}$ for space-time derivatives $\frac{\partial}{\partial x^{\mu}}$.}
\begin{eqnarray}
&&Grad:h^{(s)}(x;a)\Rightarrow (Grad\, h)^{(s+1)}(x;a) = (a\nabla)h^{(s)}(x;a)\,, \label{0.2}\\
&&Div:h^{(s)}(x;a)\Rightarrow (Div\, h)^{(s-1)}(x;a) = \frac{1}{s}(\nabla\partial_{a})h^{(s)}(x;a)\,,\label{0.3}\\
&&Tr:h^{(s)}(x;a)\Rightarrow (Tr\, h)^{(s-2)}(x;a) = \frac{1}{s(s-1)}\Box_{a}h^{(s)}(x;a)\,.\label{0.4}
\end{eqnarray}
Moreover we introduce the notation $*_a, *_b,\dots$ for a full contraction of $s$ symmetric indices:
\begin{eqnarray}
  *^{(s)}_{a}&=&\frac{1}{(s!)^{2}} \prod^{s}_{i=1}\overleftarrow{\partial}^{\mu_{i}}_{a}\overrightarrow{\partial}_{\mu_{i}}^{a} .
   \label{0.5}
\end{eqnarray}
These operators, together with their duals\footnote{It is easy to see that the operators $(a\partial_{b}), a^{2}, b^{2}$ are dual (or adjoint) to $(b\partial_{a}),\Box_{a},\Box_{b}$ with respect to the ``star'' product of tensors with two sets of symmetrized indices  (\ref{0.5})
\begin{eqnarray}
    &&\frac{1}{n}(a\partial_{b})f^{(m-1,n)}(a,b)*_{a,b} g^{(m,n-1)}(a,b)= f^{(m-1,n)}(a,b)*_{a,b} \frac{1}{m}(b\partial_{a})g^{(m,n-1)}(a,b) ,\nonumber\quad\quad\\
   && a^{2}f^{(m-2,n)}(a,b)*_{a,b} g^{(m,n)}(a,b)=f^{(m-2,n)}(a,b)*_{a,b} \frac{1}{m(m-1)}\Box_{a} g^{(m,n)}(a,b) . \quad\quad\nonumber
\end{eqnarray}
In the same fashion gradients and divergences are dual with respect to the full scalar product in the space $(x,a,b)$, where we allow for integration by parts:
\begin{eqnarray}
  (a\nabla)f^{(m-1,n)}(x;a,b)*_{a,b} g^{(m,n)}(x;a,b) &=& -f^{(m-1,n)}(x;a,b)*_{a,b}\frac{1}{m}(\nabla\partial_{a}) g^{(m,n)}(x;a,b) .\nonumber
  \end{eqnarray}
Analogous equations can be formulated for the operators $b^{2}$ or $b\nabla$.} 
will be the building blocks of the correlation functions of higher spin currents. As it was mentioned before, we use the formulation of \cite{Osborn:1993cr} reviewed in Appendix \ref{A}. Here we just extend this formulation of the two-point correlation function for the case of general spin-$s$ conformal conserved (traceless-transverse) currents.

First of all, we construct the traceless projector for rank $s$ symmetric tensors:
\begin{align}\label{0.6}
  T^{(s)}_{traceless}(a)&=\E^{(s)}(a,b)*^{(s)}_{b}T^{(s)}(b)
\end{align}
 Starting from the ansatz
\begin{eqnarray}
  && \E^{(s)}(a,b)=\sum^{s/2}_{p=0}\lambda_{p}(ab)^{s-2p}(a^{2}b^{2})^{p} , \quad \lambda_{0}=1\label{0.7}
\end{eqnarray}
and solving the tracelessness condition
\begin{eqnarray}
  && \Box_{a}\E^{(s)}(a,b)=  \Box_{b}\E^{(s)}(a,b)=0\label{0.8}
\end{eqnarray}
we arrive at a set of coefficients $\{\lambda_{p}\}^{s/2}_{p=0}$ which are the object of the recursion equation:
\begin{eqnarray}
  && \lambda_{p}=-\frac{(s-2p+2)(s-2p+1)}{4p(d/2+s-p-1)} \lambda_{p-1}\label{0.9}
\end{eqnarray}
with solution corresponding to the initial condition from (\ref{0.7}):
\begin{eqnarray}
  &&  \lambda_{p}=\frac{(-1)^{p}[s]_{2p}}{2^{2p}p![d/2+s-2]_{p}}\label{0.10}
\end{eqnarray}
Here we use notations $[a]_{n}$ for falling factorials (Phochhammer symbols):
\begin{eqnarray}
  && [a]_{n}=\frac{a!}{(a-n)!}=\frac{\Gamma(a+1)}{\Gamma(a-n+1)}\label{0.11}
\end{eqnarray}
Then it is easy to construct spin $s$ representation for inversion matrix given by:
\begin{equation}\label{0.12}
 I(a,b;x)=(ab)-2(a\hx)(b\hx),  \quad \hx_{\mu}=\frac{x_{\mu}}{\sqrt{x^{2}}}
\end{equation}
To do that we just take the traceless part of the $s$-th power of the inversion matrix:
\begin{eqnarray}
  \I^{(s)}(a,b;x)&=&\big(I(a,c;x)\big)^{s} *^{s}_{c}\E^{(s)}(c,b)=\E^{(s)}(a,c)*^{s}_{c}(I(c,b;x))^{s}\label{0.13}\\
  \Box_{a,b} \I^{(s)}(a,b;x)&=&0\label{0.14}
\end{eqnarray}
The result is easy to handle
\begin{eqnarray}\label{0.15}
  && \I^{(s)}(a,b;x)=\sum^{s/2}_{p=0}\lambda_{p}\big(I(a,b;x)\big)^{s-2p}(a^{2}b^{2})^{p}\,, \quad \lambda_{0}=1\,.
\end{eqnarray}
Then we search for two point function of conformal conserved currents with spin $s$:
\begin{eqnarray}
  && \J^{(s)}(a;x)=\J^{(s)}_{\mu_{1}\mu_{2}\dots \mu_{s}}(x) a^{\mu_{1}}a^{\mu_{2}}\dots a^{\mu_{s}} \label{0.16}\\
  && (\nabla\partial_{a})\J^{(s)}(a;x)=0 \label{0.17}\\
  && \Box_{a} \J^{(s)}(a;x)=0 \label{0.18}
\end{eqnarray}

The natural proposal is
\begin{equation}\label{0.19}
  \big<\J^{(s)}(a;x_{1}) \J^{(s)}(b;x_{2}) \big>=\frac{C_{\J}}{(x_{12}^{2})^{\Delta_{(s)}}}\I^{(s)}(a,b;x_{12})
\end{equation}
This expression is traceless by construction due to (\ref{0.14}). The scaling number $\Delta_{(s)}$ we can obtain from conservation condition (\ref{0.17}) applied to (\ref{0.19}) :
\begin{eqnarray}
  &&0= (\nabla_{1}\partial_{a})\frac{\I^{(s)}(a,b;x_{12})}{(x_{12}^{2})^{\Delta_{(s)}}}\nonumber \\
  &&= \frac{2(\Delta_{(s)}-s-d+2)}{(x_{12}^{2})^{\Delta_{(s)}+1}}
   \sum_{k=0}^{s/2-1}\lambda_{k}(s-2k)\big(I(a,b;x_{12})\big)^{s-2k-1}(b\hx_{12}) (a^{2}b^{2})^{k}\nonumber\\\label{0.20}
\end{eqnarray}
So we see that we should choose for the conformal dimension of spin $s$ field standard value:
\begin{eqnarray}
  && \Delta_{(s)}=s+d-2 \label{0.21}
\end{eqnarray}
Equivalently we can say that the conservation of the two-point function (\ref{0.19}) comes from the following relation :
\begin{align}\label{0.22}
  &\big[(\nabla_{x}\partial_{a})-2\frac{(\hx\partial_{a})}{\sqrt{x^{2}}}\big]\I^{(s)}(a,b;x)
\end{align}
The interesting point here is that if we start with expression (\ref{0.19}), where we take the correct conformal dimension (\ref{0.21}) but in expression (\ref{0.15})  undefined general set of coefficients $\lambda_{k}$ then after implementation of conservation condition we arrive to the same recursion (\ref{0.15}) for set $\lambda_{k}$ which we obtained before from the tracelessness condition (\ref{0.9}) or equivalently (\ref{0.14}).

For the odd spin case, the generalization is straightforward: we should just replace $s/2$ in summation limit by integer part $[s/2]$, which means that the highest trace, in this case, produces a vector instead of a scalar.

\section{Three-point function: the structure of the ansatz}
\label{Ansatz}
\setcounter{equation}{0}
\quad For the construction of the three-point function we should investigate structure, symmetry, and conservation condition for object $t^{j_1 j_2 i_3} (X)$, which lives in three different representations of different spins but depends locally from one point in space-time (see \cite{Osborn:1993cr} or Appendix \ref{A} for details). New important restrictions on the correlators enter the game for conserved currents: the corresponding conservation conditions should be implemented independently, restricting the correlators further. These we consider in the next section. 

First note that restricting our structure to the
 \begin{align}
t^{i_1 i_2 i_3}(X)=t^{i_1 i_2 i_3}(\hX),\label{1.1}
 \end{align}
where
\begin{align}
\hX_{\mu}=\frac{X_{\mu}}{\sqrt{X^{2}}}\,,\qquad X_{12\mu} = -X_{21\mu} = \frac{x_{13\mu}}{x_{13}^{\, 2}} -
\frac{x_{23\mu}}{x_{23}^{\, 2}} \,,\label{1.2}
\end{align}
is unit vector,  we have $q=0$ in (\ref{A.6}) and (\ref{A.8})-(\ref{A.10}). 
Taking into account that the nonsingular, tensorial part of the two-point function is given by the inversion matrix which is a function of the same unit vector (\ref{A.11}), we see from (\ref{A.5}), (\ref{A.6}) that the scaling behavior of conformal correlators depends on dimensions of fields only.

Now we formulate a general three-point function for the case of the correlation functions of three different higher-spin traceless currents.
Rewriting the (\ref{A.5}) for different spins $s_{1}, s_{2}, s_{3}$,  we get:
\begin{eqnarray}
&&\langle\J^{(s_{1})}(a;x_{1}) \,\J^{(s_{2})}(b;x_2) \, \J^{(s_{3})}(c;x_3) \r = \nonumber\\
&&=\frac{\I^{(s_{1})}(a,a';x_{13}) \I^{(s_{2})}(b,b';x_{23})*^{(s_{1})}_{a'}*^{(s_{2})}_{b'}t^{(s_{3})}(a',b';c;\hX_{12}) }{ x_{12}^{\Delta_{(s_{1})}+\Delta_{(s_{2})}-\Delta_{(s_{3})}} x_{23}^{\Delta_{(s_{2})}+\Delta_{(s_{3})}-\Delta_{(s_{1})}} x_{31}^{\Delta_{(s_{1})}+\Delta_{(s_{3})}-\Delta_{(s_{2})}}}\label{1.3}
\end{eqnarray}
where for $t^{(s_{3})}(a,b;c;\hX_{12})$ we should propose a general ansatz. For that we note that this object is traceless in all three sets of symmetrized indices, therefore we can define it as a  ``kernel'' object $\tilde{t}^{(s_{3})}(a,b;c;\hX)$ enveloped by three traceless projectors
\begin{equation}\label{1.4}
t^{(s_{3})}(\ta,\tb;\tc;\hX)=\E^{(s_{1})}(\ta,a)*_{a}\E^{(s_{2})}(\tb,b)*_{b}\tilde{t}^{(s_{3})}(a,b;c;\hX)*_{c}\E^{(s_{3})}(c,\tc)
\end{equation}
Then for $\tilde{t}^{(s_{3})}(a,b;c;\hX)$ we propose the following ansatz:
\begin{align}
 &\tilde{t}^{(s_{3})}(a,b;c;\hX) = I^{s_{3}}(c,c';\hX)*_{c'}\tH(a,b,c';\hX)\label{1.5}
\end{align}
where
\begin{align}\label{1.6}
  \tH(a,b,c;\hX)&= \sum_{\substack{\ell_{1}, \ell_{2}, \ell_{3}\in \mathcal{A}}}\tC_{\ell_{1}\ell_{2}\ell_{3}}(\hX a)^{\ell_{1}}(\hX b)^{\ell_{2}}(\hX c)^{\ell_{3}}(ab)^{\alpha}(bc)^{\beta}(ca)^{\gamma}
\end{align}
To define scope of indices $\mathcal{A}$ we  note that natural restriction:
\begin{align}
  &\alpha +\gamma + \ell_{1}=s_{1} \nonumber\\
   &\alpha +\beta + \ell_{2}=s_{2}\nonumber\\
   &\gamma +\beta + \ell_{3}=s_{3}\label{1.7}
\end{align}
completely fix $\alpha, \beta, \gamma$ for any choice of  $\ell_{1},\ell_{2},\ell_{3}$:
\begin{align}
  2\alpha&=s_{1}+s_{2}-s_{3} +\ell_{3}-\ell_{1}-\ell_{2}\label{1.8}\\
  2\beta&=s_{2}+s_{3}-s_{1} +\ell_{1}-\ell_{2}-\ell_{3}\label{1.9}\\
  2\gamma&=s_{1}+s_{3}-s_{2} +\ell_{2}-\ell_{1}-\ell_{3}\label{1.10}\\
  2(\alpha&+\beta+\gamma)=\sum s_{i}-\sum\ell_{i}\label{1.11}
\end{align}
So introducing:
\begin{align}\label{1.12}
  n_{i}=s_{i}-\ell_{i}, \quad i=1,2,3.
\end{align}
we have:
\begin{align}
  2\alpha&=n_{1}+n_{2}-n_{3}\label{1.13}\\
  2\beta&=n_{2}+n_{3}-n_{1}\label{1.14}\\
  2\gamma&=n_{1}+n_{3}-n_{2}\label{1.15}
\end{align}
and therefore from positiveness  of $\alpha,\beta,\gamma$   we have triangle inequalities:
\begin{align}
  &n_{i}+n_{j}\geq n_{k}, \quad i\neq j\neq k.\label{1.16}
\end{align}
These inequalities completely fix the scope of $\ell_{i}$ and define the number of nonzero independent parameters in our ansatz (\ref{1.6}). For general conformal dimensions of our currents, these are the only restrictions on the number of structures. The short representations, corresponding to (partially-)conserved currents, will be discussed later.

We analyzed the inequalities given above for arbitrary triplets of spins and were able to guess the analytical expressions for the number of terms in the ansatz. Interestingly, this number is not a smooth function of spins, which manifests itself by gaps when some spins coincide and different dependence of even and odd spins. We will use the step function in the following:
\begin{align}\label{1.17}
 \eta(s)=\frac{1-(-1)^{s}}{2}
\end{align}
Then the solution for numbers of allowed monomials  in the case when all spins are the same $s_{1}=s_{2}=s_{3}=s$ is
\begin{align}\label{1.18}
  &N_{sss}=\frac{1}{24}(s+2-\eta(s))(s+3)(s+4+\eta(s))
\end{align}
Then we turn to the case when two out of three spins are equal. There is a special point in this case: $s_{1}=s_{2}=s, s_{3}=2s$. The number of structures in this case is:
\begin{align}\label{1.19}
 N_{ss2s}=\frac{1}{6}(s+1)(s+2)(s+3)
 \end{align}
There are two cases beyond this point: 
\begin{itemize}
  \item $s_{3}> s=s_{1}=s_{2}$
  \begin{align}\label{1.20}
    &N^{s_{3}>s}_{sss_{3}}=\frac{1}{6}(s+1)(s+2)(s+3)-\frac{1}{24}p(p+2)(p+4)-\frac{1}{8}(p+2)\eta(p)
  \end{align}
  where $p=2s-s_{3}$, and
  \item $s_{1}<s=s_{2}=s_{3}$
  \begin{align}\label{1.21}
    & N^{s_{1}<s}_{s_{1}ss}=\frac{1}{8}[(s_{1}+2)^{2}-\eta(s_{1})](2s-s_{1}+2)
  \end{align}
\end{itemize}
Then the next observation from computer calculation is for the case $s_{1}+s_{2}=s_{3}$:
\begin{align}\label{1.22}
 N^{s_{1}+s_{2}=s_{3}}_{s_{1}s_{2}s_{3}}=\frac{1}{2}(s_{1}+1)(s_{1}+2)(s_{2}-\frac{1}{3}(s_{1}-3))
\end{align}
And finally the last observation is about numbers of monomials for the case with just general ordering $s_{1}< s_{2}< s_{3}$:
\begin{align}\label{1.23}
 &N^{s_{1}<s_{2}<s_{3}}_{s_{1}s_{2}s_{3}}= N^{s_{1}+s_{2}=s_{3}}_{s_{1}s_{2}s_{3}}-\frac{1}{24} P (P+2) (2 P+5)-\frac{1}{8}\eta(P)\\
 &P=s_{1}+s_{2}-s_{3}\nonumber
 \end{align}
So we see that (\ref{1.18})-(\ref{1.23}) completely cover all scope of indices $\mathcal{A}$ and we have analytic formula for number of  all monomials in our ansatz with indices satisfying  triangle inequalities.  
The last question remains, what happens when in our different spin case the greatest one stops to satisfy triangle inequality $s_3 > s_1 +s_2$ ? The answer is that number of monomials in this case stabilized with latest one satisfying triangle inequality $ N^{s_{1}+s_{2}<s_{3}}_{s_{1}s_{2}s_{3}}= N^{s_{1}+s_{2}=s_{3}}_{s_{1}s_{2}s_{3}}$.

Finalizing this consideration we present some geometric arguments for cubic behaviour and discontinuities in points with coincident spins.
Let us rewrite our inequality (\ref{1.16}) in the form of equations introducing three new nonnegative variables $\lambda_{i}$ 
\begin{align}
  &n_{i}+n_{j}= n_{k}+\lambda_{k}, \quad i\neq j\neq k\label{1.24}
\end{align}
then summing any pair of these equations we come to the important relation:
\begin{align}
 &\lambda_{i}+\lambda_{j}= 2n_{k}, \quad \quad i\neq j\neq k\label{1.25}\\
 &n_{i}\in [0, 1, \dots s_{i}]\label{1.26}
\end{align}
So replacing r.h.s. with maximal value we see that the scope of allowed indices is integer numbers with the following restrictions:
\begin{itemize}
  \item From (\ref{1.25}) we see that allowed $\lambda_{i}$ are all even or odd, so we have separate even or odd lattice.
  \item these even or odd pairs restricted by positiveness and inequality
  \begin{align}
   &\lambda_{i}+\lambda_{j} \le 2s_{k} \quad \quad i\neq j\neq k .\label{1.27}
  \end{align}
\end{itemize}
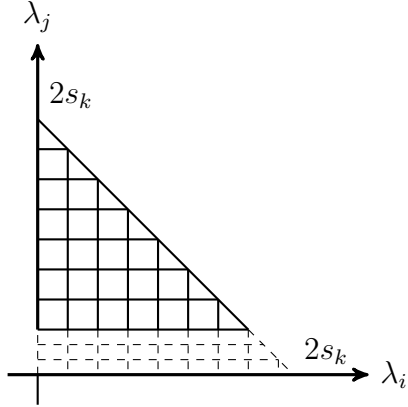
\begin{figure}[h]
\begin{center}
\begin{tikzpicture}[
    scale=4,
    axis/.style={very thick, ->, >=stealth'},
    important line/.style={thick},
    dashed line/.style={dashed, thin},
    pile/.style={thick, ->, >=stealth', shorten <=2pt, shorten
    >=2pt},
    every node/.style={color=black}
    ]
    \draw[axis] (-0.1,0)  -- (1.1,0) node(xline)[right]
        {$\lambda_{i}$};
    \draw[axis] (-0.0,.15) -- (0,1.1) node(yline)[above] {$\lambda_{j}$};
    \draw[dashed line] (0,0) -- (0,1.1); 
     \draw[important line] (0,-0.1) -- (0,0); 
    \draw[important line] (-0.0,.85)node[above right]{$2s_{k}$} coordinate (C) -- (.70,.15);
        \draw[dashed line] (.70,.15) -- (.85,-0.0)coordinate (D) node[above right, text width=5em] {$2s_{k}$}; 
        \draw[important line] (-0.0,.75) -- (.10,.75);
        \draw[important line] (-0.0,.65) -- (.20,.65);
        \draw[important line] (-0.0,.55) -- (.30,.55);
        \draw[important line] (-0.0,.45) -- (.40,.45);
        \draw[important line] (-0.0,.35) -- (.50,.35);
        \draw[important line] (-0.0,.25) -- (.60,.25);
        \draw[important line] (-0.0,.15) -- (.70,.15);
        \draw[dashed line] (-0.0,.10) -- (.75,.10);
        \draw[dashed line] (-0.0,.05) -- (.80,.05);
        \draw[important line] (.10,.75) -- (.10,.15);
        \draw[important line] (.20,.65) -- (.20,.15); 
        \draw[important line] (.30,.55) -- (.30,.15); 
        \draw[important line] (.40,.45) -- (.40,.15); 
        \draw[important line] (.50,.35) -- (.50,.15); 
        \draw[important line] (.60,.25) -- (.60,.15);
        \draw[dashed line] (.10,.15) -- (.10,-0.0);
        \draw[dashed line] (.20,.15) -- (.20,-0.0);
        \draw[dashed line] (.30,.15) -- (.30,-0.0); 
        \draw[dashed line] (.40,.15) -- (.40,-0.0); 
        \draw[dashed line] (.50,.15) -- (.50,-0.0); 
        \draw[dashed line] (.60,.15) -- (.60,-0.0);  
        \draw[dashed line] (.70,.15) -- (.70,-0.0); 
        \draw[dashed line] (.80,.05) -- (.80,-0.0);        
\end{tikzpicture}
\end{center}
\caption{Area of $\lambda_{i}+\lambda_{j} \le 2s_{k}$}
\end{figure}   

The allowed points occupy all integer vertexes of the lattice triangle in Figure 1. So the number of these points should be proportional to the area of this triangle.
To get the general picture of the numbers of allowed monomials in our ansatz, we should expand our discrete triangle in the third direction in the form of a triangle prism with a hight in the third direction. Then the full solution will be intersection of three different  prisms constructed on planes $(\lambda_{1}, \lambda_{2})$ , $(\lambda_{2}, \lambda_{3})$ and $(\lambda_{3}, \lambda_{1})$ with corresponding legs $2s_{3}, 2s_{1}, 2s_{2}$ of right triangle bases (See  Figure 2). This picture explains everything about the non-smooth behavior of our formulas above because of an irregular intersection of these prisms for different spins $s_{1}, s_{2}, s_{3}$.
\begin{figure}[h]
\centering
\includegraphics[scale=0.75]{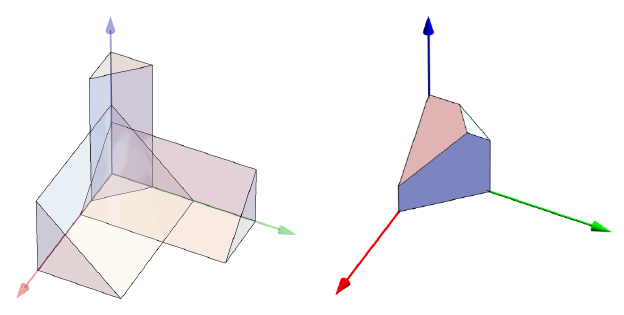}
\caption{Intersection of Prisms in the case   $s_{1}\le s_{2}\le s_{3}$}
\label{F1}
\end{figure}   

Then we can understand that in coincident cases the geometrical figures we get as a result of intersections of our prisms are more symmetric. We illustrate this for the cases $s_{1}=s_{2}\le s_{3}$ and  $s_{1}\le s_{2}= s_{3}$ (see Figure 3) and the most symmetric case $s_{1}=s_{2}=s_{3}$ (Figure 4).
\begin{figure}[h]
\includegraphics[scale=0.75]{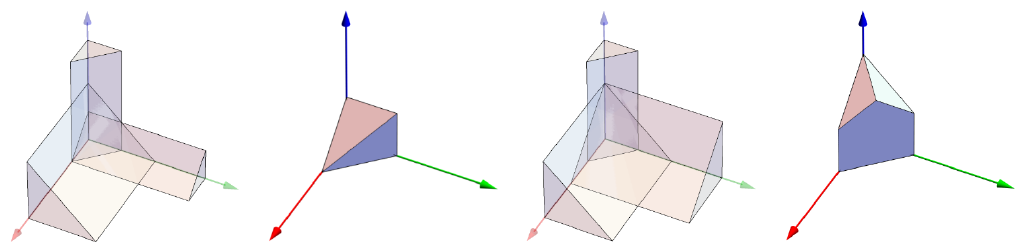}
\caption{Intersection in the case $s_{1}=s_{2}\le s_{3}$ and $s_{1}\le s_{2}=s_{3}$}
\end{figure}  
\begin{figure}[h]
\centering
\includegraphics[scale=0.75]{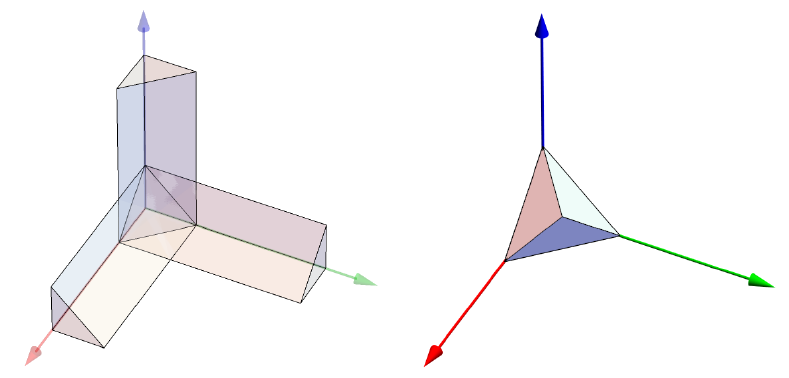}
\caption{Intersection of Prisms in the case   $s_{1}=s_{2}=s_{3}$}
\end{figure}

So we see that something like ``phase transitions'' happen in our formulas.  On the other hand, this geometrical three-dimensional picture with previous consideration of Figure 1 leads to the understanding that the full number of monomials allowed by triangle inequalities is proportional to the volume of our intersection and therefore should be a {\it cubic function of spins}.

We also note that the number of correlators for three different spins, given by equation \eqref{1.23}, coincides with the counting of \cite{Costa:2011mg}. However, in the coincident spin cases, we have a different number of correlators. The reason is that our ansatz is (anti)symmetric when exchanging any two coincident spins. Therefore, we count only the correlators with (anti)symmetric Chan-Paton factors for currents with coinciding (odd) even spin. For the case, when the currents of coincident spin have also coincident conformal weight, the correlator is indeed (anti)symmetric, therefore our ansatz covers all possible correlators. In particular, for all correlators with only conserved currents, our ansatz is a good starting point to impose the conservation conditions. This same observation can be made at the level of vertices in the bulk dual: cubic vertices of massless fields are (anti)symmetric under the exchange of fields with coinciding spins. The number of conformal correlators for unconstrained currents of spins $s_1, s_2, s_3$ we computed here is the number of symmetric traceless $SO(d)$ tensors in the product of three symmetric traceless $SO(d)$ tensors of ranks $s_1, s_2, s_3$, as known from \cite{Kravchuk:2016qvl}. For coincident spins, however, one would need to compute the symmetric product of the corresponding tensor representations.

In the end, we note that all the examples considered in \cite{Osborn:1993cr} can be exactly produced from our general formulas (\ref{1.4})-(\ref{1.6}) with corresponding choice of the value of spins and solution of the triangle inequality. 
For illustration, we discuss the important case of coinciding spins in Appendix B.

\section{Three-point function: conservation condition}
\setcounter{equation}{0} 
Now we turn to the investigation of the conservation condition for higher spin three point function.
To formulate it for higher spin case we first introduce  short notation for combinations of dimensions:
\begin{eqnarray}
  \Delta_{12} &=& \Delta_{(s_{1})}+\Delta_{(s_{2})}-\Delta_{(s_{3})} \label{2.1}\\
   \Delta_{23} &=& \Delta_{(s_{2})}+\Delta_{(s_{3})}-\Delta_{(s_{1})}\label{2.2}\\
   \Delta_{31} &=& \Delta_{(s_{3})}+\Delta_{(s_{1})}-\Delta_{(s_{2})}\label{2.3}\\
   \Delta_{(s_{i})}&=& d+s_{i}-2 ,\quad\quad i=1,2,3 \label{2.4}
\end{eqnarray}
The latter expressions are the dimensions of conserved currents.
Then redirecting readers for details of derivation to the last part of Appendix A,  we  can write conservation condition 
\begin{align}\label{2.5}
 & (\nabla_{x_{1}}\partial_{a})\langle\J^{(s_{1})}(a;x_{1}) \,\J^{(s_{2})}(b;x_2) \, \J^{(s_{3})}(c;x_3) \r =0
\end{align}
in the form:
\begin{eqnarray}
  (\nabla_{X}\partial_{a})t^{(s_{3})}(a,b;c;X) &=\Delta_{12}\frac{(X\partial_{a})}{X^{2}}t^{(s_{3})}(a,b;c;X) \label{2.6}
\end{eqnarray}
The last one is the equation for structural tensor object $t^{(s_{3})}(a,b;c;X)$ which is completely equivalent to the conservation condition for the three-point function.
Then, separating the traceless projector from the ``kernel'' part of (\ref{1.4}) (see also Appendix A for details) and introducing 
the $k-$th trace of our ansatz:
\begin{align}
  &\square^{k}_{a}\tilde{t}^{(s)}(a,b,c;\hX)=  \sum_{\substack{\ell_{1}\in [2k,\dots s_{1}];\ell_{2},\ell_{3}\in [0,\dots s_{2},s_{3}]\\\{\ell_{i}\}\in \A}}T^{(k)}_{\ell_{1},\ell_{2},\ell_{3}}\gff{\ell_{1}-2k,\ell_{2},\ell_{3}}{\alpha;\beta,\gamma}\label{2.7}
\end{align}
where we shortened the formulas using the notation:
\begin{align}
  &\gff{\ell_{1},\ell_{2},\ell_{3}}{\alpha;\beta,\gamma}=(\hX a)^{\ell_{1}}(\hX b)^{\ell_{2}}(\hX c)^{\ell_{3}}(ab)^{\alpha}I^{\beta}(b,c;\hX)I^{\gamma}(c,a;\hX)\label{2.8} 
\end{align}
and $T^{(k)}_{\ell_{1},\ell_{2},\ell_{3}}$ is $k-$th trace map of $\tC_{\ell_{1}\ell_{2}\ell_{3}}$ from (\ref{1.6}).
In this way using important formula (\ref{A.30}) and expression (\ref{2.7}) after long manipulations we write conservation condition (\ref{2.6}) in terms of equations on $T^{(k)}_{\ell_{1},\ell_{2},\ell_{3}}$:
\begin{align}
&(\ell_{1}-2k)(s_{3}-s_{2})T^{(k)}_{\ell_{1},\ell_{2},\ell_{3}}\nonumber\\
  +&(\alpha+1)(2\ell_{3}-2k-d-2s_{2}+2)T^{(k)}_{\ell_{1}-1,\ell_{2}-1,\ell_{3}}
  +(\gamma+1)(2\ell_{2}-2k-d-2s_{3}+2)T^{(k)}_{\ell_{1}-1,\ell_{2},\ell_{3}-1}\nonumber\\
  +&(\alpha+1)(\ell_{3}+1)T^{(k)}_{\ell_{1}-1,\ell_{2},\ell_{3}+1}
  +(\gamma+1)(\ell_{2}+1)T^{(k)}_{\ell_{1}-1,\ell_{2}+1,\ell_{3}}\nonumber\\
  +&\frac{1}{d+2s_{1}-2k-4}\left[2(\ell_{2}-\ell_{3})T^{(k+1)}_{\ell_{1},\ell_{2},\ell_{3}}
  +2(\beta+1)\big(T^{(k+1)}_{\ell_{1}+1,\ell_{2},\ell_{3}-1}+T^{(k+1)}_{\ell_{1}+1,\ell_{2}-1,\ell_{3}}\big)\right.\nonumber\\
  -&\left.(\ell_{2}+1)T^{(k+1)}_{\ell_{1}+1,\ell_{2}+1,\ell_{3}}-(\ell_{3}+1)T^{(k+1)}_{\ell_{1}+1,\ell_{2},\ell_{3}+1}\right]=0 \label{2.9}
  \end{align}
where the traces themselves  satisfy the following recursion relation:
\begin{gather}
T^{(k+1)}_{\ell_{1},\ell_{2},\ell_{3}}=(\ell_{1}-2k)(\ell_{1}-2k-1)T^{(k)}_{\ell_{1},\ell_{2},\ell_{3}}
+2(\alpha+1)(\gamma+1)T^{(k)}_{\ell_{1}-2,\ell_{2},\ell_{3}}\nonumber\\
+2(\alpha+1)(\ell_{1}-2k-1)T^{(k)}_{\ell_{1}-1,\ell_{2}-1,\ell_{3}}-2(\gamma+1)(\ell_{1}-2k-1)T^{(k)}_{\ell_{1}-1,\ell_{2},\ell_{3}-1}\label{2.10}
\end{gather}
That is not the whole story.
The bad news here is that the equation (\ref{2.9}) should be supplemented by a conservation condition for the second current in the correlation function when the latter is also conserved. This can be done in (\ref{2.6}) by replacements of $s_{1}\leftrightarrow s_{2}$ and $ x_{1} \leftrightarrow x_{2}$ and $a_{\mu}\leftrightarrow b_{\mu}$, or directly in 
(\ref{2.9}), (\ref{2.10}) replacing $s_{1}\leftrightarrow s_{2}, \ell_{1}\leftrightarrow \ell_{2}$.

The good news here is that we do not need to solve all recursion equations (\ref{2.9}) for all $T^{(k)}_{\ell_{1},\ell_{2},\ell_{3}}\, (k=0,1\dots [s_{1}/2])$.  
In fact, we need to solve only the first conservation condition for $k=0$, all others will be satisfied automatically because they are higher ($k-$th) traces of the first one with $k=0$.

Using the helpful ansatz-normalization:
\begin{eqnarray}
 && T^{(0)}_{\ell_{1},\ell_{2},\ell_{3}}= \frac{(-1)^{\ell_{3}}}{\alpha!\beta!\gamma!}C_{\ell_{1},\ell_{2},\ell_{3}}\label{2.11}\\
  &&T^{(1)}_{\ell_{1},\ell_{2},\ell_{3}}=\frac{(-1)^{\ell_{3}}}{\alpha!\beta!\gamma!}\Big[\ell_{1}(\ell_{1}-1)C_{\ell_{1},\ell_{2},\ell_{3}}
+2\beta C_{\ell_{1}-2,\ell_{2},\ell_{3}}\nonumber\\
&&+2(\ell_{1}-1)C_{\ell_{1}-1,\ell_{2}-1,\ell_{3}}+2(\ell_{1}-1)C_{\ell_{1}-1,\ell_{2},\ell_{3}-1}\Big]= \frac{(-1)^{\ell_{3}}}{\alpha!\beta!\gamma!} T_{\ell_{1},\ell_{2},\ell_{3}} \label{2.12}
\end{eqnarray}
we obtain effective conservation condition: 
\begin{align}
&\ell_{1}(s_{3}-s_{2})C_{\ell_{1},\ell_{2},\ell_{3}}\nonumber\\
  +&(2\ell_{3}-d-2s_{2}+2)C_{\ell_{1}-1,\ell_{2}-1,\ell_{3}}
  -(2\ell_{2}-d-2s_{3}+2)C_{\ell_{1}-1,\ell_{2},\ell_{3}-1}\nonumber\\
  +&(\ell_{2}+1)C_{\ell_{1}-1,\ell_{2}+1,\ell_{3}}
  -(\ell_{3}+1)C_{\ell_{1}-1,\ell_{2},\ell_{3}+1}\nonumber\\
  +&\frac{1}{d+2s_{1}-4}\left[2(\ell_{2}-\ell_{3})T_{\ell_{1},\ell_{2},\ell_{3}}
  +2(\beta+1)\big(T_{\ell_{1}+1,\ell_{2},\ell_{3}-1}+T_{\ell_{1}+1,\ell_{2}-1,\ell_{3}}\big)\right.\nonumber\\
  -&\left.(\ell_{2}+1)T_{\ell_{1}+1,\ell_{2}+1,\ell_{3}}-(\ell_{3}+1)T_{\ell_{1}+1,\ell_{2},\ell_{3}+1}\right]=0 \label{2.13}
  \end{align}
which we should amend with the same type of equation but now for $s_2$, if the second current is also conserved:
\begin{align}
&\ell_{2}(s_{3}-s_{1})C_{\ell_{1},\ell_{2},\ell_{3}}\nonumber\\
  +&(2\ell_{3}-d-2s_{1}+2)C_{\ell_{1}-1,\ell_{2}-1,\ell_{3}}
  -(2\ell_{1}-d-2s_{3}+2)C_{\ell_{1},\ell_{2}-1,\ell_{3}-1}\nonumber\\
  +&(\ell_{1}+1)C_{\ell_{1}+1,\ell_{2}-1,\ell_{3}}
  -(\ell_{3}+1)C_{\ell_{1},\ell_{2}-1,\ell_{3}+1}\nonumber\\
  +&\frac{1}{d+2s_{2}-4}\left[2(\ell_{1}-\ell_{3})\bar{T}_{\ell_{1},\ell_{2},\ell_{3}}
  +2(\gamma+1)\big(\bar{T}_{\ell_{1},\ell_{2}+1,\ell_{3}-1}+\bar{T}_{\ell_{1}-1,\ell_{2}+1,\ell_{3}}\big)\right.\nonumber\\
  -&\left.(\ell_{1}+1)\bar{T}_{\ell_{1}+1,\ell_{2}+1,\ell_{3}}-(\ell_{3}+1)\bar{T}_{\ell_{1},\ell_{2}+1,\ell_{3}+1}\right]=0 \label{2.14}
  \end{align}
where $T_{\ell_{1},\ell_{2},\ell_{3}}, \bar{T}_{\ell_{1},\ell_{2},\ell_{3}}$ are corresponding trace maps:
\begin{align}
  &T_{\ell_{1},\ell_{2},\ell_{3}}= \Big[\ell_{1}(\ell_{1}-1)C_{\ell_{1},\ell_{2},\ell_{3}}
+2\beta C_{\ell_{1}-2,\ell_{2},\ell_{3}}\nonumber\\
&+2(\ell_{1}-1)C_{\ell_{1}-1,\ell_{2}-1,\ell_{3}}+2(\ell_{1}-1)C_{\ell_{1}-1,\ell_{2},\ell_{3}-1}\Big] \label{2.15}\\
&\bar{T}_{\ell_{1},\ell_{2},\ell_{3}}=\Big[\ell_{2}(\ell_{2}-1)C_{\ell_{1},\ell_{2},\ell_{3}}
+2\gamma C_{\ell_{1},\ell_{2}-2,\ell_{3}}\nonumber\\
&+2(\ell_{2}-1)C_{\ell_{1}-1,\ell_{2}-1,\ell_{3}}+2(\ell_{2}-1)C_{\ell_{1},\ell_{2}-1,\ell_{3}-1}\Big] \label{2.16}
\end{align}
We do not yet have a full solution for this system of equations. But we analyzed these equations using a computer program and investigated the rank of this linear system for different triplets of spins using our ansatz (\ref{1.4})-(\ref{1.6}) and normalization (\ref{2.11}), (\ref{2.12}). This system of linear equations for $C_{\ell_{1},\ell_{2},\ell_{3}}$ has a number of independent parameters satisfying triangle inequalities described in the previous section. Then computing the rank of the corresponding system for multiple cases we obtain a universal answer: the rank of the system (\ref{2.13}), (\ref{2.14}) depends only on the minimal spin:
\begin{itemize}
  \item {\it The number of independent parameters of the three-point function (or linearly independent correlators) of conserved currents with spins $s_1,s_2,s_3$ is equal to}
  $$N_{s_1,s_2,s_3}=min\{s_1,s_2,s_3\}+1\,.$$
\end{itemize}
We refer to Appendix B for some details on the special case of coincident spins. 

\section{Conservation condition as a differential equation}
\setcounter{equation}{0}

In this section, we first construct differential equations for the correlators of conserved currents in the case of coincident spins and then generalize them to the cases with different spins. 
First, we transform our recursion equation (\ref{B.20}) to a differential equation multiplying it by the following powers of formal variables $x^{\ell_{1}-1} y^{\ell_{2}} z^{\ell_{3}}$ and summing on all possible values of $\ell^{i}$
\begin{align}
  D(\partial_{x},\partial_{y},\partial_{z};C(x,y,z))&=\sum_{\{\ell_{i}\}}D_{\ell_{1}\ell_{2}\ell_{3}} x^{\ell_{1}-1} y^{\ell_{2}} z^{\ell_{3}}=0\label{3.1}
\end{align}
 in other words we should obtain differential equation for the functions
\begin{equation}
  C(x,y,z)=\sum_{\{\ell_{i}\}}C_{\ell_{1}\ell_{2}\ell_{3}}x^{\ell_{1}} y^{\ell_{2}} z^{\ell_{3}}\label{3.2}
\end{equation}
and
\begin{equation}
  T(x,y,z)=\sum_{\{\ell_{i}\}}T_{\ell_{1}\ell_{2}\ell_{3}}x^{\ell_{1}-2} y^{\ell_{2}} z^{\ell_{3}}\label{3.3}
\end{equation}
In all these equations $\{\ell_{i}\}$ means value of indeces $\ell_{i}, i=1,2,3$ satisfying the triangle inequality
\begin{equation}
  s+\ell_{i}\geq \ell_{j}+\ell_{k},\quad i\ne j\ne k \label{3.4}
\end{equation}
Comparing (\ref{3.3}) with (\ref{B.21})we obtain:
\begin{align}
 &T(x,y,z)=[\partial^{2}_{x} + (x+2y+2z)\partial_{x}-y\partial_{y}-z\partial_{z}+s+2]C(x,y,z)\label{3.5}
\end{align}
Then we can obtain differential equation version of our conservation equation (\ref{B.20}):
\begin{align}
  &D(\partial_{x},\partial_{y},\partial_{z};C(x,y,z))\nonumber\\
  &=\Big[(\Delta_{s}+s)(z-y)
  +\frac{1}{2}(s+1-4yz+x\partial_{x}-y\partial_{y}-z\partial_{z})(\partial_{y}-\partial_{z})\Big]C(x,y,z)\nonumber\\
  &+\frac{1}{d+2s-4}\Big[(2x+y+z+\frac{1}{2}[\partial_{y}+\partial_{z}])(y\partial_{y}-z\partial_{z})\nonumber\\
  &+(s-x\partial_{x})(y-z-\frac{1}{2}[\partial_{y}-\partial_{z}])\Big]T(x,y,z)=0\label{3.6}
\end{align}
We see that our differential operator is antisymmetric in $z$ and $y$ although the functions $C(x,y,z)$ and $T(x.y.z)$ are symmetric.
A generalization to different spins is straightforward: instead of (\ref{3.6}) we have an equation obtained with the same scheme from the recursion equation (\ref{2.13}):
\begin{align}
  &D^{(s_{1},s_{2},s_{3})}(\partial;C(x,y,z))= \big[(s_{3}-s_{2})\partial_{x}+(\Delta_{s_{3}}+s_{3})z-(\Delta_{s_{2}}+s_{2})y\big]C(x,y,z)\nonumber\\
  &+\frac{1}{2}(s_{2}+s_{3}-s_{1}+1-4yz+x\partial_{x}-y\partial_{y}-z\partial_{z})(\partial_{y}-\partial_{z})C(x,y,z)\nonumber\\
  &+\frac{1}{d+2s_{1}-4}\Big[(2x+y+z+\frac{1}{2}[\partial_{y}+\partial_{z}])(y\partial_{y}-z\partial_{z})\nonumber\\
  &+(s_{1}-x\partial_{x})(y-z-\frac{1}{2}[\partial_{y}-\partial_{z}])+\frac{1}{2}(s_{3}-s_{2})[y+z+\partial_{y}+\partial_{z}]\Big]T(x,y,z)=0\nonumber\\\label{3.7}
\end{align}
where $T(x,y,z)$ in this case is
\begin{align}
 &T(x,y,z)=[\partial^{2}_{x} + (x+2y+2z)\partial_{x}-y\partial_{y}-z\partial_{z}+s_{2}+s_{3}-s_{1}+2]C(x,y,z)\quad \label{3.8}
\end{align}
The equation (\ref{3.7}) should be supplemented by a conservation condition for the second current, when the latter is conserved. This can be obtained from (\ref{3.7}) and (\ref{3.8}) by replacements $s_{1}\leftrightarrow s_{2}$ and $ x \leftrightarrow y$. The solution to these general equations for the correlators of conserved currents will be addressed in an upcoming work.

\section{Conclusions}

We have established a general ansatz for the tensorial structure of the conformal three-point function for general spins and general dimensions. This allows us to calculate the exact numbers of conformal structures corresponding to all cases of AdS dual bulk interaction vertices. We present explicit formulas for three-point functions of conformal correlators of three non-conserved currents, corresponding to massive fields in the bulk. The number of structures for non-conserved currents is equivalent to the number of vertices with massive fields in the bulk, counting the number of contractions of three symmetric fields of ranks $s_1,s_2,s_3$ with each other and derivatives acting on them, with a condition that the traces and divergences are excluded, and the derivatives do not contract between themselves (this latter condition, stemming from field-redefinition freedom, limits the possible Lorentz scalars to a finite number: see, e.g., \cite{Metsaev:2005ar,Joung:2012rv,Kessel:2018ugi}).
For coincident spins, our counting for correlators, symmetric under the exchange of the coinciding spin currents, (\ref{1.18})-(\ref{1.22}) are new to our best knowledge. For all different spins, there cannot be symmetry under exchange of currents, thus our counting \eqref{1.23} coincides with that of \cite{Costa:2011mg}.

The special cases of (partially) conserved currents, corresponding to the short representations or (partially-)massless fields in the bulk, will be studied elsewhere: the extra constraints on the correlators stemming from the conservation of the currents imply non-trivial differential equations, for which the general solutions will be treated in future work. However, we worked out and further studied the structure of the constraints in the case of the conserved currents, both as differential equations and as recursion relations on the coefficients of the ansatz. The latter form allowed us to tackle a large number of cases numerically. Our results confirm the expectation from earlier works \cite{Costa:2011mg,Costa:2011dw,Stanev:2012nq,Zhiboedov:2012bm} about the number of structures in the correlators of conserved currents, which, in turn, coincides with the number of massless vertices in the bulk \cite{Metsaev:2005ar,Manvelyan:2010jr}.

The conservation condition comes with technical subtleties as the operator of the divergence imposing the conservation of the currents in the ansatz does not commute with the traceless projector. Our careful treatment takes into account the trace terms in the projector properly.

We hope to solve analytically the conservation conditions to fully classify the correlators of (partially-)conserved currents and make a match with the $AdS$ vertices involving (partially-)massless fields \cite{Joung:2012hz}. The case of all massive fields is fully covered by our ansatz in one-to-one correspondence with the vertices in the bulk \cite{Metsaev:2005ar,Joung:2012rv}, assuming symmetry under exchange of the currents/fields of coincident spins.

The correlation functions of three conserved currents were derived earlier using different approaches in \cite{Stanev:2012nq,Zhiboedov:2012bm}. In even dimensions, they were described by the correlators in free theories of so-called singletons --- conformal fields describing the short conformal representations described by the (self-dual) multi-forms, corresponding to rectangular Young diagrams of the half-maximal height of the massless little group in even dimensions (see, e.g., \cite{Bekaert:2009fg}). In four dimensions, these are the spin-s massless fields, which are representations of the conformal algebra $SO(4,2)$ despite the lack of conformal symmetry in their standard off-shell descriptions (see, e.g., \cite{Joung:2014qya,Barnich:2015tma}).\footnote{
Explicit descriptions of the singleton theories in terms of covariant Lagrangians are so far only well-studied for the spin-one case (see \cite{Evnin:2022kqn} for a review).
} The situation is different in the odd dimensions \cite{Zhiboedov:2012bm}, where the singletons are missing or, presumably, correspond to some generalized free field theories lacking locality: free field equations containing square root of d'Alambertian operator (see, e.g., \cite{Joung:2015jza}).

The formulation \cite{Osborn:1993cr} and our generalization for higher spins are also suitable for the investigation of the singular part of the correlation function to get a route to the trace anomaly structure in the higher-spin case. We leave this to future investigations.

\section*{Acknowledgements}
R. M. would like to thank Stefan Theisen, Rubik Poghossian and Aleksey Isaev for many valuable discussions during long period of preparation of this paper, and special gratitude to Ruben Mkrtchyan for productive  and focused on result discussions.
R. M. and M. K. where supported by the Science Committee of RA, in the frames of
the research project \# 21AG-1C060.
K. M. was supported by the
European Union’s Horizon 2020 Research and Innovation
Programme under the Marie Sk\l odowska-Curie Grant
No.\ 844265, UKRI and STFC Consolidated Grants ST/T000791/1 and ST/X000575/1.

\appendix

\section{Short review of Osborn-Petkou formulation and adaptation to higher spin case}
\label{A}

\renewcommand{\theequation}{A.\arabic{equation}}\setcounter{equation}{0}
In this appendix, we present a short review of useful formulas and constructions proposed in article \cite{Osborn:1993cr} (see also \cite{Osborn:1994rv,Erdmenger:1996yc}).
\subsection*{Conformal Transformations}
The conformal transformations (combination of translation, rotation, scale transformation, and special conformal boosts)  are diffeomorphisms preserving metric up to a local scale factor:
\begin{eqnarray}
  && x_{\mu} \rightarrow x'_{\mu}(x), \quad g_{\mu\nu} dx'^{\mu}dx'^{\nu} \rightarrow \Omega(x)^{-2} g_{\mu\nu} dx^{\mu}dx^{\nu}\label{A.1}
\end{eqnarray}
Combining this transformation with local dilatation we arrive at local rotations:
\begin{eqnarray}
  && R_{\mu}^{\;\alpha}(x)=\Omega(x)\frac{\partial x'_{\mu}}{\partial x_{\alpha}} ,\quad R_{\mu}^{\;\alpha}(x) R_{\alpha}^{\;\nu}(x)=\delta_{\mu}^{\nu}\,. \label{A.2}
\end{eqnarray}
Adding inversion to this picture :
\begin{eqnarray}
  && x'_{\mu}=\frac{x_{\mu}}{x^{2}}, \quad \Omega(x)=x^{2},\quad R_{\mu\nu}(x)=I_{\mu\nu}(x)=\delta_{\mu\nu}-2\frac{x_{\mu}x_{\nu}}{x^{2}}\label{A.3}
\end{eqnarray}
we see that the rotation operator, in this case, is the Inversion matrix $I_{\mu\nu}$. 
A combination of inversion, rotation, and translation can describe any conformal transformation.

We will show below how the conformal symmetry fixes the form of the two and three-point correlation functions for arbitrary quasi-primary fields $\mathcal{O}^{i}(x)$, where $i$ is an index counting corresponding representation of the rotation group $O(d)$ (see \cite{Osborn:1993cr} for details). The symmetric representation of the conformal group is defined by two quantum numbers: the spin and the conformal dimension. The two-point function of two operators is fixed by conformal symmetry up to an overall constant:
 \begin{eqnarray}
   && <\mathcal{O}^{i}(x_{1})\mathcal{\bar{O}}_{j}(x_{2})>=\frac{C_{\mathcal{O}}}{(x_{12}^{2})^{\eta}} D^{i}_{j}(I(x_{12})), \quad x_{12\mu}=x_{1\mu}-x_{2\mu}\label{A.4}
 \end{eqnarray}
Here $\mathcal{\bar{O}}_{j}(x)$ is conjugate representation for $\mathcal{O}^{i}(x)$ with the same conformal dimension. Another important object here is $D(I(x_{12}))$ which is corresponding representation for the inversion matrix $I_{\mu\nu}(x)=\delta_{\mu\nu}-2x_{\mu}x_{\nu}/x^{2}$ .
\subsection*{Three point function}
Since conformal transformations transform any three points into any others, the three-point function is also essentially defined
in general dimension $d$. Our discussion for arbitrary representations
for the fields $\O_1 , \O_2 , \O_3$ with dimensions $\eta_{1}, \eta_{2}, \eta_{3}$ is based on the following formula from \cite{Osborn:1993cr}
\begin{align}
\langle \O_1^{i_1}(x_1)\, \O_2^{i_2} (x_2) \, \O_3^{i_3} (x_3) \r &=
\frac{1}{(x_{12}^{\, 2})^{\de_{12}}\,(x_{23}^{\, 2})^{\de_{23}}\,
(x_{31}^{\, 2})^{\de_{31}}}\nonumber\\
& \times D_1^{\, i_1} {}_{\! j_1} (I(x_{13}))
D_2^{\, i_2} {}_{\! j_2} (I(x_{23})) \, t^{j_1 j_2 i_3} (X_{12}) \, ,\label{A.5}
\end{align}
where $t^{i_1 i_2 i_3}(X)$ is a  tensor living in three different spin representations in general case. This object transforms in a proper way with respect to local rotation and dilatations.
\begin{align}
 &D_1^{\, i_1} {}_{\! j_1} (R) D_2^{\, i_2} {}_{\! j_2} (R)
D_3^{\, i_3} {}_{\! j_3} (R)\, t^{j_1 j_2 j_3} (X) =
t^{i_1 i_2 i_3}(RX) \ \hbox {for all}\ R \in O(d) \ ,\nonumber\\
&t^{i_1 i_2 i_3}(\lambda X) = {}\lambda^q t^{i_1 i_2 i_3}(X)\label{A.6}
\end{align}
and
\begin{align}
X_{12\mu} = -X_{21\mu} = \frac{x_{13\mu}}{x_{13}^{\, 2}} -
\frac{x_{23\mu}}{x_{23}^{\, 2}} \, ,
\quad X_{12}^{\, 2} = \frac{x_{12}^{\, 2}}{x_{13}^{\, 2} x_{23}^{\, 2}}\label{A.7}
\end{align}
The scaling dimensions of the fields should
satisfy the following expressions
\begin{align}
&\de_{12} =\frac{1}{2}(\eta_1 + \eta_2 - \eta_3 + q ) \, ,\label{A.8}\\
&\de_{23} =  \frac{1}{2}(\eta_2 + \eta_3 - \eta_1 - q ) \, ,\label{A.9}\\
&\de_{31} =\frac{1}{2}(\eta_3 + \eta_1 - \eta_2 - q ) \, .\label{A.10}
\end{align}
So we see that for the construction of the two-point function for spin $s$ currents, we should realize the construction of the representation of the inversion matrix $D(I(x_{12}))$ 
where:
\begin{equation}\label{A.11}
  I_{\mu\nu}(x_{12})=\delta_{\mu\nu}-2\hx_{12\mu}\hx_{12\nu}, \quad  \hx_{12}=\frac{x_{12}}{\sqrt{x^{2}_{12}}}
\end{equation}
which is more or less obvious and known.  Another important property of this formulation is that in the three-point function we can rearrange all three representations due to the following important properties \cite{Osborn:1993cr} of structural function ($q=0$):
\begin{align}
&D_1^{\, i_1} {}_{\! j_1}(I(\hx_{13}))
D_2^{\, i_2} {}_{\! j_2} (I(\hx_{23})) \, t^{j_1 j_2 i_3} (\hX_{12}) \nonumber\\
& = D_1^{\, i_1} {}_{\! j_1} (I(\hx_{12})) D_3^{\, i_3} {}_{\! j_3} (I(\hx_{32})) \, {\tilde t}^{\, j_1 i_2 j_3}
(\hX_{13}) = D_2^{\, i_2} {}_{\! j_2} (I(\hx_{21}))
D_3^{\, i_3} {}_{\! j_3} (I(\hx_{31})) \, {\hat t}^{\, i_1 j_2 j_3} (\hX_{32}) \, , \nonumber\\
& {\tilde t}^{\, i_1 i_2 i_3} (\hX) =  D_1^{\, i_1} {}_{\! j_1}(I(\hX)) \, t^{j_1 i_2 i_3} (\hX)  , \quad {\hat t}^{\, i_1 i_2 i_3} (\hX)
=  D_2^{\, i_2} {}_{\! j_2} (I(\hX)) \, t^{i_1 j_2 i_3} (\hX) \, . \label{A.12}
\end{align}
It follows then, that in the case when all three representations are the same (i.e. same spin currents) and
the three-point function is symmetric for all fields $\O_1,\, \O_2,\,
\O_3,$ then:
\begin{equation}\label{A.13}
t^{i_2 i_1 i_3}(X) = t^{i_1 i_2 i_3} (- X) \, , \quad
 D^{\, i_1} {}_{\! j_1} (I(X)) \, {t}^{j_1 i_2 i_3} (X) =
t^{i_3 i_1 i_2} (-X).
\end{equation}
The first relation contains $-X$ in r.h.s. because this object depends on the space-time coordinates through the difference between the inversions of the first and second coordinates around the third point (\ref{A.7}), and when we replace the first two operators we also exchange $x_{1}$ with $x_{2}$.  The importance of the minus sign in the second relation we consider in detail during the investigation of our ansatz for $t^{i_1 i_2 i_3} (X)$.
Then for irreducible representations, for which the two-point functions are fixed as (\ref{A.4}), we see consistent scaling behavior and covariance with respect to inversions, rotations, and translations.
All these mean that $D(I(x_{12}))$ behaves as a parallel transport transformation between two space-time points 
for local conformal rotations. This fact is very important for understanding  
an analogous formula for three-point functions. The important property of conformal transformations is that one can map any three points into any other
three points. This leads to an essentially (almost)  unique three-point function 
in general dimension $d$. The general form of the three-point function is considered in \cite{Osborn:1993cr} and presented here in (\ref{A.5}).
The original point of this consideration is that the three-point function is described through the homogeneous tensor $t^{i_1 i_2 i_3}(X)$ satisfying
(\ref{A.6}) and (\ref{A.12}). More details can be found in \cite{Osborn:1993cr},\cite{Osborn:1994rv} and \cite{Erdmenger:1996yc}, here we just note that if we restrict ourselves to the polynomial function of unit vector:
 \begin{align}
t^{i_1 i_2 i_3}(X)=t^{i_1 i_2 i_3}(\hX),\label{A.14}
 \end{align}
where
\begin{align}
\hX_{\mu}= \frac{X_{\mu}}{\sqrt{X^{2}}},\label{A.15}
\end{align}
 then in (\ref{A.8})-(\ref{A.10}) we have
\begin{align}
  q=0 \label{A.16}
\end{align}
and instead of 
\begin{align}
I_{\mu \alpha}(x_{23}) \hX_{12\, \alpha} =\frac{x^{2}_{12} }{x^{2}_{13}} \hX_{13 \, \mu} \, , \quad
I_{\mu \alpha}(x_{13}) \hX_{12\, \alpha} = \frac{x^{2}_{12}}{x^{2}_{23}} \hX_{32 \, \mu} \, , \label{A.17}
\end{align}
 we have
\begin{align}
I_{\mu \alpha}(x_{23}) \hX_{12\, \alpha} =  \hX_{13 \, \mu} \, , \quad
I_{\mu \alpha}(x_{13}) \hX_{12\, \alpha} = \hX_{32 \, \mu} \, , \label{A.18}
\end{align}
and we see that inversion operators $I_{\mu \alpha}(x_{ij}), i\neq j, i,j=1,2,3$ really rotate from one direction to other unit inverted vectors
$\hX_{ij}$.
This leads to the familiar expression for the three-point function:
\begin{align}
\langle\O_1^{i_1}(x_1)\, \O_2^{i_2} (x_2) \, \O_3^{i_3} (x_3) \r = {}&
\frac{1}{(x_{12}^{\, 2})^{\de_{12}}\,(x_{23}^{\, 2})^{\de_{23}}\,
(x_{31}^{\, 2})^{\de_{31}}}\nonumber\\
& \times D_1^{\, i_1} {}_{\! j_1} (I(x_{13}))
D_2^{\, i_2} {}_{\! j_2} (I(x_{23})) \, t^{j_1 j_2 i_3} (\hX_{12}) \, ,\label{A.19}
\cr
\end{align}
where $t^{i_1 i_2 i_3}(\hX)$ is a homogeneous and dimensionless  tensor satisfying 
\begin{align}
 D_1^{\, i_1} {}_{\! j_1} (R) D_2^{\, i_2} {}_{\! j_2} (R)
D_3^{\, i_3} {}_{\! j_3} (R) & \, t^{j_1 j_2 j_3} (\hX) =
t^{i_1 i_2 i_3}(R\hX) \ \hbox {for all}\ R \ ,\label{A.20}\\
t^{i_1 i_2 i_3}(\lambda \hX) = {}&t^{i_1 i_2 i_3}(\hX)\label{A.21}
\end{align}
and
\begin{equation}
\hX_{12\mu} = -\hX_{21\mu} = \sqrt{\frac{x_{13}^{\, 2} x_{23}^{\, 2}}{x_{12}^{\, 2}}}\left[\frac{x_{13\mu}}{x_{13}^{\, 2}} -
\frac{x_{23\mu}}{x_{23}^{\, 2}}\right]\label{A.22}
\end{equation}
The scaling dimensions  of the fields for $q=0$ are
\begin{align}
\de_{12} = & \frac{1}{2} (\eta_1 + \eta_2 - \eta_3) \, , \nonumber\\
\de_{23} = & \frac{1}{2} (\eta_2 + \eta_3 - \eta_1) \, , \nonumber\\
\de_{31} = & \frac{1}{2} (\eta_3 + \eta_1 - \eta_2) \, . \label{A.23}
\end{align}
\subsection*{Conservation condition}
For the derivation of the conservation conditions, we note that:
\begin{align}
  & (\nabla_{x_{1}}\partial_{a})\langle\J^{(s_{1})}(a;x_{1}) \,\J^{(s_{2})}(b;x_2) \, \J^{(s_{3})}(c;x_3) \r \nonumber\\
  &=\nabla_{x_{1}^{\mu}}\left[\frac{1}{x_{12}^{\Delta_{12}}x_{23}^{\Delta_{23}} x_{31}^{\Delta_{31}}}
\partial_{a^{\mu}}\I^{(s_{1})}(a,a';x_{13})\right]\I^{(s_{2})}(b,b';x_{23})*^{(s_{1})}_{a'}*^{(s_{2})}_{b'}t^{(s_{3})}(a',b';c;\hX_{12})\nonumber\\
 &+ \frac{1}{x_{12}^{\Delta_{12}}x_{23}^{\Delta_{23}} x_{31}^{\Delta_{31}}}\partial_{a^{\mu}}\I^{(s_{1})}(a,a';x_{13})
\I^{(s_{2})}(b,b';x_{23})*^{(s_{1})}_{a'}*^{(s_{2})}_{b'}\nabla_{x_{1}^{\mu}}t^{(s_{3})}(a',b';c;\hX_{12})\label{А.24}
\end{align}
Using the following relations:
\begin{align}
&\nabla_{x_{1}^{\mu}}\frac{1}{x_{12}^{\Delta_{12}} x_{31}^{\Delta_{31}}} 
=-\frac{1}{x_{12}^{\Delta_{12}} x_{31}^{\Delta_{31}}}\left[ \frac{\Delta_{12}x_{12\mu}}{x^{2}_{12}}+\frac{\Delta_{31}x_{13\mu}}{x^{2}_{13}}\right]\nonumber\\
&=-\frac{1}{x_{12}^{\Delta_{12}} x_{31}^{\Delta_{31}}}\left[ \Delta_{12}X_{32\mu}+(\Delta_{12}+\Delta_{31})\frac{x_{13\mu}}{x^{2}_{13}}\right]\nonumber\\
&=-\frac{1}{x_{12}^{\Delta_{12}} x_{31}^{\Delta_{31}+2}}\left[ \Delta_{12}I_{\mu\alpha}(x_{13})\frac{X^{\alpha}_{12}}{X^{2}_{12}}+(\Delta_{12}+\Delta_{31})\frac{x_{13\mu}}{x^{2}_{13}}\right]\label{A.25}
\end{align}
\begin{align}
&(\nabla_{x_{1}}\partial_{a})\I^{(s_{1})}(a,a';x_{13})=2(d+s_{1}-2)\frac{(x_{13}\partial_{a})}{x_{13}^{2}}\I^{(s_{1})}(a,a';x_{13})\label{A.26}
\end{align}
\begin{align}
&\nabla_{x_{1}^{\mu}}t^{(s_{3})}(a,b;c;X_{12})
=\nabla_{X^{\alpha}_{12}}t^{(s_{3})}(a,b;c;X_{12})\frac{\partial X^{\alpha}_{12}}{\partial{x^{\mu}_{1}}}\nonumber\\
&=\nabla_{X^{\alpha}_{12}}t^{(s_{3})}(a,b;c;X_{12})\frac{I_{\mu}^{\alpha}(x_{13})}{x^{2}_{13}}\label{A.27}
\end{align}
we see that the conservation condition is satisfied when:
\begin{eqnarray}
  \Delta_{12}+\Delta_{31} &=&2\Delta_{(s_{1})}=2(d+s_{1}-2)\label{A.28}
\end{eqnarray}
and:
\begin{eqnarray}
  (\nabla_{X}\partial_{a})t^{(s_{3})}(a,b;c;X) &=\Delta_{12}\frac{(X\partial_{a})}{X^{2}}t^{(s_{3})}(a,b;c;X) \label{A.29}
\end{eqnarray}
This is the equation for structural tensor object $t^{(s_{3})}(a,b;c;X)$ which we use in the second section. The equation (\ref{A.29}) (or (\ref{2.6})) is equivalent to the conservation condition for the first current in the three-point function.

Now we can separate the traceless projector from ``kernel'' part and write (\ref{A.29}) in the following form:
\begin{align}
&\Big(\nabla^{\mu}-\Delta_{12}\frac{\hX^{\mu}}{\sqrt{X^{2}}}\Big)\partial^{a}_{\mu}\E^{(s_{1})}(a,a')*^{s_{1}}_{a'}
\tilde{t}^{(s_{3})}(a',b,c;\hX)*^{s_{2}}_{b}*^{s_{3}}_{c}\E^{(s_{2})}(b,\tb)\E^{(s_{3})}(c,\tc)\nonumber\\
&=\Scale[0.9]{{1\over s_{1}!}\sum^{s_{1}/2-1}_{k=0}(s_{1}-2k)!\lambda^{s_{1}}_{k}[a^{2}]^{k}
\Big[\Big((\nabla\partial^{a})-\Delta_{12}\frac{(\hX\partial^{a})}{\sqrt{X^{2}}}\Big)
\square^{k}_{a}}\nonumber\\
&-\Scale[0.9]{\frac{1}{d+2s_{1}-2k-4}\Big((a\nabla)-\Delta_{12}\frac{(a\hX)}{\sqrt{X^{2}}}\Big)
\square^{k+1}_{a}\Big]\tilde{t}^{(s_{3})}(a,b,c;\hX)*^{s_{2}}_{b}*^{s_{3}}_{c}\E^{(s_{2})}(b,\tb)\E^{(s_{3})}(c,\tc)}\label{A.30}
\end{align}
where $\tilde{t}^{(s_{3})}(a,b,c;\hX)$ now is:
\begin{align}
  &\tilde{t}^{(s_{3})}(a,b,c;\hX)= I^{s_{3}}(c,c';\hX)*_{c'}\tH^{(s_{123})}(a,b,c';\hX)  \nonumber\\
  & = \sum_{\substack{s_{i}\in [0,\dots s_{i}]\\ \{s_{i}\} \in \A}}(-1)^{\ell_{3}}\tC_{\ell_{1}\ell_{2}\ell_{3}}(\hX a)^{\ell_{1}}(\hX b)^{\ell_{2}}(\hX c)^{\ell_{3}}(ab)^{\alpha}I^{\beta}(b,c;\hX)I^{\gamma}(c,a;\hX).\label{A.31}
\end{align}
Then we compute the $p$-th trace as:
\begin{align}
   &\Phi^{k}(a;b,c;\hX;\alpha,\gamma)=\Box^{k}_{a}(ab)^{\alpha}(ac)^{\gamma}(\hX a)^{\ell_{1}}\nonumber\\
   &=\sum_{{p,q,n \atop p+q+n \leq k}}\rho\gf{k;p,q,n}{\alpha,\gamma,\ell_{1}}
   (ab)^{\alpha-k+n+q}(ac)^{\gamma-k+n+p}(bc)^{k-n-p-q}(\hX a)^{\ell_{1}-2n-p-q}(\hX b)^{p}(\hX c)^{q},\label{A.32}
\end{align}
where we neglected all terms of type $O(b^{2},c^{2})$.
From the equation
\begin{equation}\label{A.33}
\Phi^{k+1}(a;b,c;\hX;\alpha,\gamma)=\Box_{a}\Phi^{k}(a;b,c;\hX;\alpha,\gamma)
\end{equation}
we get the following recursion relation
\begin{align}
\rho\gf{k+1;p,q,n}{\alpha,\gamma,\ell_{1}} &=2\rho\gf{k;p,q,n}{\alpha,\gamma,\ell_{1}}(\alpha-k+n+q)(\gamma-k+n+p)\nonumber\\
&+2\rho\gf{k;p-1,q,n}{\alpha,\gamma,\ell_{1}}(\alpha-k+n+q)(\ell_{1}-2n-p-q+1)\nonumber\\
&+2\rho\gf{k;p,q-1,n}{\alpha,\gamma,\ell_{1}}(\gamma-k+n+p)(\ell_{1}-2n-p-q+1)\nonumber\\
&+\rho\gf{k;p,q,n-1}{\alpha,\gamma,\ell_{1}}(\ell_{1}-2n-p-q+2)(\ell_{1}-2n-p-q+1)\label{A.34}
\end{align}
This equation after substitution
\begin{equation}\label{A.35}
  \rho\gf{k;p,q,n}{\alpha,\gamma,\ell_{1}}=2^{k-n}[\alpha]_{k-n-q}[\gamma]_{k-n-p}[\ell_{1}]_{2n+p+q}\hat{\rho}(k;p,q,n)
\end{equation}
goes to Pascal's identity for multinomial:
\begin{align}
 \hat{\rho}(k+1;p,q,n) &=\hat{\rho}(k;p,q,n)+\hat{\rho}(k;p,q,n-1)\nonumber\\
&+\hat{\rho}(k;p-1,q,n)+\hat{\rho}(k;p,q-1,n)\label{A.36}
\end{align}
with obvious solution
\begin{equation}\label{A.37}
 \hat{\rho}(k;p,q,n)=\frac{[k]_{n+p+q}}{p!q!n!}=\binom{k}{p,q,n}
\end{equation}
Then we can easily derive the $k$th trace of our ansatz:
\begin{align}
  &\square^{k}_{a}\tilde{t}^{(s)}(a,b,c;\hX)=  \sum_{\substack{\ell_{1}\in [2k,\dots s_{1}];\ell_{2},\ell_{3}\in [0,\dots s_{2},s_{3}]\\\{\ell_{i}\}\in \A}}T^{(k)}_{\ell_{1},\ell_{2},\ell_{3}}\gff{\ell_{1}-2k,\ell_{2},\ell_{3}}{\alpha;\beta,\gamma}\label{A.38}
\end{align}
where we introduced notation:
\begin{align}
  &\gff{\ell_{1},\ell_{2},\ell_{3}}{\alpha;\beta,\gamma}=
  (\hX a)^{\ell_{1}}(\hX b)^{\ell_{2}}(\hX c)^{\ell_{3}}(ab)^{\alpha}I^{\beta}(b,c;\hX)I^{\gamma}(c,a;\hX)\label{A.39} 
\end{align}
and $T^{(k)}_{\ell_{1},\ell_{2},\ell_{3}}$ is $k$th trace map of $\tC_{\ell_{1}\ell_{2}\ell_{3}}$

\begin{align}
&T^{(k)}_{\ell_{1},\ell_{2},\ell_{3}}
=(-1)^{\ell_{3}}\sum_{{p,q,n \atop p+q+n \leq k}}\tC_{\ell_{1}-2k+2n+p+q,\ell_{2}-p,\ell_{3}-q}
\,\,\rho\gf{k;p,q,n}{\alpha,\gamma,\ell_{1}}\label{A.40}
\end{align}
In this way substituting (\ref{A.38}) in (\ref{A.30}) one can straightforwardly derive the conservation condition on $T^{(k)}_{\ell_{1},\ell_{2},\ell_{3}}$ given in (\ref{2.9}).

\section{Examples}

\renewcommand{\theequation}{B.\arabic{equation}}\setcounter{equation}{0}
\subsection*{Coincident spins $s_{1}=s_{2}=s_{3}=s$}
We present examples for the most symmetric case of equal spins $s_{1}=s_{2}=s_{3}=s$. It is enough to write a ``kernel'' term with the following symmetry properties:
\begin{align}\label{B.1}
  &\tilde{t}^{(s)}(a,b;c;\hX)=\tilde{t}^{(s)}(b,a;c;-\hX)\\
  &I^{s}(a,a';\hX)*_{a'}\tilde{t}^{(s)}(a',b;c;\hX)=\tilde{t}^{(s)}(c,a;b;-\hX)\label{B.2}
\end{align}
From these conditions, we derive the most general polynomial ansatz for $t^{(s)}(a,b;c;\hX)$:
\begin{align}
\tilde{t}^{(s)}(a,b;c;\hX)= & I^{s}(c,c';\hX)*_{c'}\tH^{(s)}(a,b,c';\hX)\label{B.3}\\
\tilde{t}_{1}^{(s)}(a,b;c;\hX) =&  \big[\tH^{(s)}(a,b,c;\hX)
   + I^{s}(a,a';\hX)*_{a'}\tH^{(s)}(a',b,c;-\hX)\nonumber\\+&I^{s}(b,b';\hX)*_{b'}\tH^{(s)}(a,b',c;\hX)\big]\label{B.4}
 \end{align}
where the main object, $\tH$, is given by
\begin{align}\label{B.5}
  \tH^{(s)}(a,b,c;\hX)&= \sum_{\substack{\ell_{1},\ell_{2},\ell_{3}\in [0,\dots s]\\ \{\ell_{i}\} \in \bar{\A}}}\tC_{\ell_{1}\ell_{2}\ell_{3}}(\hX a)^{\ell_{1}}(\hX b)^{\ell_{2}}(\hX c)^{\ell_{3}}(ab)^{\alpha}(bc)^{\beta}(ca)^{\gamma}\,.
\end{align}
Here $\bar{\A}$ is the range of indices defined by the following natural restrictions: 
\begin{align}
  &\alpha +\gamma + \ell_{1}=s \label{B.6}\\
   &\alpha +\beta + \ell_{2}=s\label{B.7}\\
   &\gamma +\beta + \ell_{3}=s\label{B.8}
\end{align}
These also can be resolved fixing  $\alpha, \beta, \gamma$ for any choice of  $\ell_{1},\ell_{2},\ell_{3}$:
\begin{align}
  2\alpha&=s +\ell_{3}-\ell_{1}-\ell_{k}\label{B.9}\\
  2\beta&=s +\ell_{1}-\ell_{2}-\ell_{3}\label{B.10}\\
  2\gamma&=s +\ell_{2}-\ell_{1}-\ell_{3}\label{B.11}
\end{align}
The positiveness of $\alpha, \beta, \gamma$ for coincident spins leads to the triangle inequalities:
\begin{align}
  s +\ell_{i}\geq \ell_{j}+\ell_{k}\, \quad i\neq j \neq k\label{B.12}
\end{align}
Another special point in consideration of equal spins is that conditions (\ref{B.1}) and (\ref{B.2}) force the coefficients $T^{(0)}_{\ell_{1}\ell_{2}\ell_{3}}$ to be completely symmetric with respect to $\ell_{1}, \ell_{2}, \ell_{3}$.
Then:
\begin{align}
  I^{s}(a,a';\hX)*_{a'}I^{s}(b,b';\hX)*_{b'}&\tH^{(s)}(a',b',c;\hX)
  =(-1)^{\ell_{1}+\ell_{2}+\ell_{3}}I^{s}(c,c';\hX)*_{c'}\tH^{(s)}(a,b,c';\hX)\nonumber\\
  &=I^{s}(c,c';\hX)*_{c'}\tH^{(s)}(a,b,c';-\hX)\label{B.13}
\end{align}
and we get even (odd) sum of $\ell$'s for even (odd) spin $s$:
\begin{equation}
 \sum_{i=1,2,3}\ell_{i}=3s-2(\alpha+\beta+\gamma)\,.\label{B.14}
\end{equation}

The relation (\ref{B.13}) helps to explain the minus sign  in condition (\ref{B.2}) and we can  make the following simple derivation showing  that (\ref{B.4}) is equivalent to (\ref{B.3}):
\begin{align}
  &\tH^{(s)}(a
,b,c;\hX) +I^{s}(a,a';\hX)*_{a'}\tH^{(s)}(a',b,c;-\hX)+I^{s}(b,b';\hX)*_{b'}\tH^{(s)}(a,b',c;\hX)\nonumber\\
&=I^{s}(c,c';\hX)*_{c'}\Big[I^{s}(c',c'';\hX)*_{c''}H^{(s)}(a
,b,c'';\hX)+I^{s}(b,b';\hX)*_{b'}\tH^{(s)}(a,b',c';\hX)\nonumber\\& +I^{s}(a,a';\hX)*_{a'}\tH^{(s)}(b ,a',c';\hX)\Big]=I^{s}(c,c';\hX)*_{c'}\bar{\tH}^{(s)}(a,b,c';\hX)\label{B.15}
\end{align}
where
\begin{align}
  \bar{\tH}^{(s)}(a,b,c';\hX) & =\sum_{\substack{\bar{\ell}_{1},\bar{\ell}_{2},\bar{\ell}_{3}\in [0,\dots s]\\\bar{\ell}_{1}+\bar{\ell}_{2}+\bar{\ell}_{3}=\textnormal{even}}}\bar{T}^{(0)}_{\bar{\ell}_{1}\bar{\ell}_{2}\bar{\ell}_{3}}(\hX a)^{\bar{\ell}_{1}}(\hX b)^{\bar{\ell}_{2}}(\hX c)^{\tilde{\ell}_{3}}(ab)^{\bar{\alpha}}(bc)^{\bar{\beta}}(ca)^{\bar{\gamma}} \,,\label{B.16}\\
  \bar{T}^{(0)}_{\bar{\ell}_{1}\bar{\ell}_{2}\bar{\ell}_{3}} &= \bar{T}^{(0)}({\bar{\ell}_{1}|\bar{\ell}_{2}\bar{\ell}_{3}})+\bar{T}^{(0)}({\bar{\ell}_{2}|\bar{\ell}_{3}\bar{\ell}_{1}})
  +\bar{T}^{(0)}({\bar{\ell}_{3}|\bar{\ell}_{1}\bar{\ell}_{2}})\,,\label{B.17}
\end{align}
where (symmetric in all $\bar{\ell}_{i}; i=1,2,3$) coefficients $\bar{T}^{(0)}_{\bar{\ell}_{1}\bar{\ell}_{2}\bar{\ell}_{3}}$ are constructed as a cyclic permutation (\ref{B.17}) of the object that is symmetric in two indices only:
\begin{align}
   \bar{T}^{(0)}({\bar{\ell}_{1}|\bar{\ell}_{2}\bar{\ell}_{3}})& =(-1)^{\bar{\ell}_{1}}\sum_{n_{2},n_{3}}^{\bar{\ell}_{2},\bar{\ell}_{3}}
   2^{n_{2}+n_{3}}T^{(0)}_{\bar{\ell}_{1}-n_{2}-n_{3},\bar{\ell}_{2}-n_{2}\bar{\ell}_{3}-n_{3}}
   \binom{\bar{\alpha}+n_{2}}{\bar{\alpha}}\binom{\bar{\gamma}+n_{3}}{\bar{\gamma}}\label{B.18}
\end{align}
The most general ansatz in this case is (\ref{B.3}), with traceless projectors written as:
\begin{equation}
t^{(s)}(\ta,\tb;\tc;\hX)=\E^{(s)}(\ta,a)*_{a}\tilde{t}^{(s)}(a,b;c;\hX)*_{b}*_{c}\E^{(s)}(b,\tb)\E^{(s)}(c,\tc).\label{B.19}
\end{equation}
\subsection*{Conservation condition for coincident spins}
When all spins coincide, we need only one equation for fully symmetric coefficients:
\begin{align}
  &(\alpha+1)(2\ell_{3}-2k-\Delta_{(s)}-s)T^{(k)}_{\ell_{1}-1,\ell_{2}-1,\ell_{3}}
  +(\gamma+1)(2\ell_{2}-2k-\Delta_{(s)}-s)T^{(k)}_{\ell_{1}-1,\ell_{2},\ell_{3}-1}\nonumber\\
  +&(\alpha+1)(\ell_{3}+1)T^{(k)}_{\ell_{1}-1,\ell_{2},\ell_{3}+1}
  +(\gamma+1)(\ell_{2}+1)T^{(k)}_{\ell_{1}-1,\ell_{2}+1,\ell_{3}}\nonumber\\
  +&\frac{1}{d+2s-2k-4}\left[2(\ell_{2}-\ell_{3})T^{(k+1)}_{\ell_{1},\ell_{2},\ell_{3}}
  +2(\beta+1)\big(T^{(k+1)}_{\ell_{1}+1,\ell_{2},\ell_{3}-1}+T^{(k+1)}_{\ell_{1}+1,\ell_{2}-1,\ell_{3}}\big)\right.\nonumber\\
  -&\left.(\ell_{2}+1)T^{(k+1)}_{\ell_{1}+1,\ell_{2}+1,\ell_{3}}-(\ell_{3}+1)T^{(k+1)}_{\ell_{1}+1,\ell_{2},\ell_{3}+1}\right]=0\label{B.20}
  \end{align}
where
\begin{gather}
T^{(k+1)}_{\ell_{1},\ell_{2},\ell_{3}}=(\ell_{1}-2k)(\ell_{1}-2k-1)T^{(k)}_{\ell_{1},\ell_{2},\ell_{3}}
+2(\alpha+1)(\gamma+1)T^{(k)}_{\ell_{1}-2,\ell_{2},\ell_{3}}\nonumber\\
+2(\alpha+1)(\ell_{1}-2k-1)T^{(k)}_{\ell_{1}-1,\ell_{2}-1,\ell_{3}}-2(\gamma+1)(\ell_{1}-2k-1)T^{(k)}_{\ell_{1}-1,\ell_{2},\ell_{3}-1}\,,\label{B.21}
\end{gather}
and we need to solve only the first conservation condition for $k=0$ (the rest follow from tracelessness). Using the helpful ansatz (\ref{2.11}), (\ref{2.12}) we arrive to the following recursion for (symmetric in $\ell_{1},\ell_{2},\ell_{3}$) expressions $C_{\ell_{1},\ell_{2},\ell_{3}}$ and $T_{\ell_{1},\ell_{2},\ell_{3}}$
\begin{align}
  &D_{\ell_{1},\ell_{2},\ell_{3}}=(2\ell_{3}-\Delta_{(s)}-s)C_{\ell_{1}-1,\ell_{2}-1,\ell_{3}}
  -(2\ell_{2}-\Delta_{(s)}-s)C_{\ell_{1}-1,\ell_{2},\ell_{3}-1}\nonumber\\
  +&\beta(\ell_{2}+1)C_{\ell_{1}-1,\ell_{2}+1,\ell_{3}}-\beta(\ell_{3}+1)C_{\ell_{1}-1,\ell_{2},\ell_{3}+1}
  \nonumber\\
  +&\frac{1}{d+2s-4}\left[2(\ell_{2}-\ell_{3})T_{\ell_{1},\ell_{2},\ell_{3}}
  +2\big(\gamma T_{\ell_{1}+1,\ell_{2}-1,\ell_{3}}-\alpha T_{\ell_{1}+1,\ell_{2},\ell_{3}-1}\big)\right.\nonumber\\
  +&\left.\gamma(\ell_{3}+1)T_{\ell_{1}+1,\ell_{2},\ell_{3}+1}-\alpha(\ell_{2}+1)T_{\ell_{1}+1,\ell_{2}+1,\ell_{3}}\right]=0\label{B.22}
  \end{align}
where
\begin{gather}
T_{\ell_{1},\ell_{2},\ell_{3}}=\ell_{1}(\ell_{1}-1)C_{\ell_{1},\ell_{2},\ell_{3}}
+2\beta C_{\ell_{1}-2,\ell_{2},\ell_{3}}\nonumber\\
+2(\ell_{1}-1)C_{\ell_{1}-1,\ell_{2}-1,\ell_{3}}+2(\ell_{1}-1)C_{\ell_{1}-1,\ell_{2},\ell_{3}-1}\label{B.23}
\end{gather}
Computer-assisted solutions have $s+1$ independent parameters as they should.

\subsection*{Spin 2 case: energy-momentum tensor and connection with (\ref{B.3}), (\ref{B.5})}

First we review construction in the case of spin two following \cite{Osborn:1993cr}.
For three point function of energy-momentum tensors we have:
\begin{align}
\langle T_{\mu \nu}(x_1) \,T_{\si\rho} (x_2) \, T_{\alpha \beta} (x_3) \r
= {}&{1\over x_{12}^{\, d \vphantom \eta}\,
x_{13}^{\, d \vphantom \eta}\,x_{23}^{\, d \vphantom \eta}} \,
\I_{\mu\nu,\mu'\nu'}(x_{13}) \I_{\si\rho,\si'\rho'}(x_{23}) \,
t_{\mu'\nu'\si'\rho'\alpha \beta} (X_{12}) \, , \label{B.24}
\end{align}
with $t_{\mu\nu\si\rho\alpha\beta}(X)$ homogeneous of degree zero
in $X$, symmetric and traceless on
each pair of indices $\mu\nu, \ \si\rho$ and $\alpha\beta$ and from
satisfying
\begin{align}
&t_{\mu\nu\si\rho\alpha\beta}(X) = {}
t_{\si\rho\mu\nu\alpha\beta}(X) \, .\label{B.25}\\
&\I_{\mu\nu,\mu'\nu'}(X)t_{\mu'\nu'\si\rho\alpha\beta}(X)={}
t_{\alpha\beta\mu\nu\si\rho}(X) \, ,\label{B.26}
\end{align}
The conservation equations require just
\begin{equation}
 \Bigl ( \pr_\mu - d\, {X_\mu \over X^2} \Bigl )
t_{\mu\nu\si\rho\alpha\beta}(X) = 0 \label{B.27}
\end{equation}
Defining
\begin{eqnarray}
h^1_{\mu \nu}(\hX) &=& \hX_\mu \hX_\nu - {1\over d} \, \de_{\mu \nu}
\, , \quad \hX_\mu = {X_\mu \over \sqrt {X^2}}
\label{B.28}\\
 h^2_{\mu \nu \si \rho}(\hX) &=& {}\hX_\mu \hX_{\si}\de_{\nu \rho}
+ (\mu\leftrightarrow\nu,\si\leftrightarrow\rho)\nonumber\\
&-&{4\over d}
\hX_\mu \hX_\nu\de_{\si \rho} -{4\over d} \hX_\si \hX_\rho \de_{\mu\nu}
+ {4\over d^2} \de_{\mu \nu} \de_{\si\rho} \qquad
\label{B.29}\\
h^3_{\mu \nu \si \rho} &=& {} \de_{\mu \si} \de_{\nu\rho}
+\de_{\mu \rho} \de_{\nu\si} -
{2\over d}\, \de_{\mu \nu} \de_{\si\rho}=2\E_{\mu\nu,\si\rho}\label{B.30}\\
 h^4_{\mu\nu\si\rho\alpha\beta}(\hX)&=& {} h^3_{\mu\nu\si\alpha}
\hX_\rho \hX_\beta + (\si\leftrightarrow\rho,\alpha\leftrightarrow
\beta)\nonumber\\
&-& {2\over d} \, \de_{\si \rho}^{\vphantom h}h^2_{\mu\nu\alpha\beta}(\hX)
- {2\over d} \, \de_{\alpha\beta}^{\vphantom h}h^2_{\mu\nu\si\rho}(\hX)
-{8\over d^2}\, \de_{\si\rho}^{\vphantom h}\de_{\alpha\beta}^{\vphantom h}
h^1_{\mu\nu}(\hX) \, ,\label{B.31}\\
h^5_{\mu\nu\si\rho\alpha\beta} &=& {} \de_{\mu\si}^{\vphantom h}
\de_{\nu\alpha}^{\vphantom h}\de_{\rho\beta}^{\vphantom h}
+ (\mu\leftrightarrow\nu,\si\leftrightarrow\rho,
\alpha\leftrightarrow\beta)\nonumber\\
&-&{4\over d}\,\de_{\mu\nu}^{\vphantom h} h^3_{\si\rho\alpha\beta}
-{4\over d}\,\de_{\si\rho}^{\vphantom h} h^3_{\mu\nu\alpha\beta}
-{4\over d}\,\de_{\alpha\beta}^{\vphantom h} h^3_{\mu\nu\si\rho}
-{8\over d^2} \, \de_{\mu\nu}^{\vphantom h}\de_{\si\rho}^{\vphantom h}
\de_{\alpha\beta}^{\vphantom h}\,,\label{B.32}
\end{eqnarray}
a general expansion for $t_{\mu\nu\si\rho\alpha\beta}(X)$
has the form
\begin{align}
t_{\mu\nu\si\rho\alpha\beta}(X) = {}& a\,
h^5_{\mu\nu\si\rho\alpha\beta} + b\,
h^4_{\alpha\beta\mu\nu\si\rho}(\hX) + b'\bigl (
h^4_{\mu\nu\si\rho\alpha\beta}(\hX) +
h^4_{\si\rho\mu\nu\alpha\beta}(\hX)\bigl) \cr
& + c\, h^3_{\mu\nu\si\rho}h^1_{\alpha\beta}(\hX) + c'\bigl (
h^3_{\si\rho\alpha\beta}h^1_{\mu\nu}(\hX) +
h^3_{\mu\nu\alpha\beta}h^1_{\si\rho}(\hX)\bigl) \cr
& + e\, h^2_{\mu\nu\si\rho}(\hX)h^1_{\alpha\beta}(\hX) + e'\bigl (
h^2_{\si\rho\alpha\beta}(\hX)h^1_{\mu\nu}(\hX) +
h^2_{\mu\nu\alpha\beta}(\hX)h^1_{\si\rho}(\hX)\bigl) \cr
& +f\, h^1_{\mu\nu}(\hX)h^1_{\si\rho}(\hX)h^1_{\alpha\beta}(\hX) \,.\label{B.33}
\end{align}
From the symmetry condition (\ref{B.25}), (\ref{B.26}) we have
\begin{align}
b+b'=-2a\, , \quad c'=c \, , \quad e+e'=-4b'-2c \, ,\label{B.34}
\end{align}
so that $a,b,c,e,f$ may be regarded as independent. Then using conservation condition (\ref{B.23}) we have two additional constraints:
\begin{align}
d^2 a + 2(b+b') -(d-2)b'-dc +e'={}& 0 \, , \label{B.35}\\
d(d+2)(2b'+c)+4(e+e')+f={}& 0 \, . \label{B.36}
\end{align}
Therefore, we have three undetermined independent coefficients, say, $a,b,c$, which are the free parameters of the three-point function (in arbitrary dimension $d$):
\begin{align}
f=(d+4)(d-2)(4a+2b-c),\label{B.37}\\
 e'=-(d+4)(d-2)a-(d-2)b+dc,\label{B.38}\\
 e=(d+2)(da+b-c).\label{B.39}
\end{align}
Now we can compare these with our general formulation in the case of spin two.
We should look at ansatz (\ref{B.3}) and (\ref{B.5}) for the case of $s=2$. First of all putting $s=2$ in corresponding number of solution of triangle inequality (\ref{1.18}) we obtain
$N_{222}=5$ which is correct number of parameters after applying symmetry constraints (\ref{B.3}) then investigating these independent five terms in ansatz (\ref{B.5}), identifying with (\ref{B.33}) and using notation
\begin{equation}\label{B.40}
  \tC_{\ell_{1},\ell_{2},\ell_{3}}=(-1)^{\ell_{3}}C_{\ell_{1},\ell_{2},\ell_{3}}
\end{equation}
where $C_{\ell_{1},\ell_{2},\ell_{3}}$ is symmetric in $\ell_{1},\ell_{2},\ell_{3}$.
we obtain the following connections between coefficients:
\begin{align}
  & a={C_{000}\over 8}; \quad b={C_{110}\over 8}; \quad b'=-{C_{000} \over 4}-{C_{110} \over 8}; \label{B.41}\\
  & c=c'={C_{200}\over 2}; \quad e=C_{000}+C_{110}+{C_{112}\over 4};\label{B.42}\\
  & e'=-{C_{110} \over 2}-{C_{112} \over 4}-C_{200}; \quad f=4C_{110}+4C_{112}+8C_{200}+C_{222}.\label{B.43}
\end{align}
So we see that these 8 coefficients $a,b,b',c,c',e,e',f$ from \cite{Osborn:1993cr} expressed through the five coefficients from our ansatz $C_{000}, C_{110}, C_{200}, C_{112}, C_{222}$. Because triangle inequality and symmetricity of $C_{\ell_{1},\ell_{2},\ell_{3}}$  lead to the solution (\ref{B.25}), (\ref{B.26}) in general case.
Then we can investigate conservation condition (\ref{B.27}). taking into account that our normalization here slightly differ and we should insert in (\ref{B.5})
\begin{equation}\label{B.44}
 \tC_{\ell_{1},\ell_{2},\ell_{3}}=\alpha!\beta!\gamma!C_{\ell_{1},\ell_{2},\ell_{3}}
\end{equation}
we see that for $s=2$ we have only two nonzero independent equations:
\begin{align}
  &D_{1,1,0}\sim (8-d^{2}-2d)C_{000} +(6-d)C_{110}+(4d+8)C_{200}+2C_{112}=0\label{B.45}\\
  &D_{1,2,1}\sim (d^{2}-12)C_{110}-12C_{112}-2d(d-2)C_{200}-4C_{222}=0\label{B.46}
\end{align}
Now we see that it is possible to express $C_{112}$ and $C_{222}$ through  the  remaining three arbitrary parameter $C_{000}$, $C_{110}$ and $C_{200}$ and these free parameters from (\ref{B.45}), (\ref{B.46}) are exactly equivalent to $a,b,c$ (see (\ref{B.41}) and (\ref{B.42})). Moreover after some straightforward manipulation we can see that all relations (\ref{B.34})-(\ref{B.39}) are also satisfied exactly.  
\subsection*{Spin 3 case: solution of the conservation condition (\ref{B.22})}
Finalizing this Appendix we just present solution of the conservation condition for spin three case. Here we have  eight different parameters in our ansatz and conservation equations expressed four from them through the four independent :
\begin{align}
&C_{3,0,0}=\frac{1}{9 (d+2)}\Big[(d-2) (d+8) C_{1,0,0}+(d-14) C_{2,1,0}-2 C_{1,1,1}-2 C_{2,2,1}\Big]\\
&C_{3,1,1}=\frac{1}{6 (d+2)}\Big[(d+8) (d-2)^2 C_{1,0,0}+(d (d+2)+8) C_{1,1,1}\nonumber\\
&\quad\quad\quad\quad\quad\quad\quad\quad\quad\quad\quad\quad-4 (d (d+8)-4) C_{2,1,0}+8 C_{2,2,1}\Big]\\
&C_{3,2,2}=\frac{1}{12 (d+2)}\Big[-(d+6) (d+8) (d-2)^2 C_{1,0,0}-2 (d (d+10)+32) C_{1,1,1}\nonumber\\
&\hspace{2.5cm}+ 2 (d^{3}+24d^{2}+60d-96) C_{2,1,0}+2 (d(d-12)-44) C_{2,2,1}\Big]\\
&C_{3,3,3}=\frac{1}{54 (d+2)}\Big[-(d+8)(d-2)^2 (d^{2}-10d-60) C_{1,0,0}\nonumber\\
&+(640-d^{4}+2d^{3}-12d^{2}-200d) C_{1,1,1}+4 (d^{4}+3d^{3}-124d^{2}-300d+480) C_{2,1,0}\nonumber\\
&+(3 d^{3}-16d^{2}+180d+736) C_{2,2,1}\Big]
\end{align}

\end{document}